\documentclass[journal]{IEEEtran}

\usepackage[font=small]{caption}
\usepackage{graphicx,graphics,epsfig,epstopdf,float}
\usepackage{amsmath} 
\usepackage{amssymb}  
\usepackage{mathrsfs}
\usepackage{enumerate}
\usepackage{subfigure}
\usepackage{colortbl}
\usepackage{color}  
\usepackage{bm}
\usepackage{psfrag}
\usepackage{cite}
\usepackage{algorithm}
\usepackage{algorithmic}
\usepackage{mdwlist}
\usepackage{longtable}
\usepackage{array}
\usepackage{acronym}  

\usepackage{pifont}
\usepackage{graphicx}
\usepackage{subfig}
\usepackage{bbm}
\usepackage{extarrows}
\usepackage{cleveref}
\usepackage{zref-xr}
\zxrsetup{toltxlabel=true, tozreflabel=false}

\DeclareMathAlphabet{\mathsfbr}{OT1}{cmss}{m}{n}
\SetMathAlphabet{\mathsfbr}{bold}{OT1}{cmss}{bx}{n}
\DeclareRobustCommand{\msf}[1]{%
  \ifcat\noexpand#1\relax\msfgreek{#1}\else\mathsfbr{#1}\fi
}
\makeatletter
\newcommand{\msfgreek}[1]{\csname s\expandafter\@gobble\string#1\endcsname}
\makeatother

\DeclareFontEncoding{LGR}{}{} 
\DeclareSymbolFont{sfgreek}{LGR}{cmss}{m}{n}
\SetSymbolFont{sfgreek}{bold}{LGR}{cmss}{bx}{n}
\DeclareMathSymbol{\salpha}{\mathord}{sfgreek}{`a}
\DeclareMathSymbol{\sbeta}{\mathord}{sfgreek}{`b}
\DeclareMathSymbol{\sgamma}{\mathord}{sfgreek}{`g}
\DeclareMathSymbol{\sdelta}{\mathord}{sfgreek}{`d}
\DeclareMathSymbol{\sepsilon}{\mathord}{sfgreek}{`e}
\DeclareMathSymbol{\szeta}{\mathord}{sfgreek}{`z}
\DeclareMathSymbol{\seta}{\mathord}{sfgreek}{`h}
\DeclareMathSymbol{\stheta}{\mathord}{sfgreek}{`j}
\DeclareMathSymbol{\siota}{\mathord}{sfgreek}{`i}
\DeclareMathSymbol{\skappa}{\mathord}{sfgreek}{`k}
\DeclareMathSymbol{\slambda}{\mathord}{sfgreek}{`l}
\DeclareMathSymbol{\smu}{\mathord}{sfgreek}{`m}
\DeclareMathSymbol{\snu}{\mathord}{sfgreek}{`n}
\DeclareMathSymbol{\sxi}{\mathord}{sfgreek}{`x}
\DeclareMathSymbol{\somicron}{\mathord}{sfgreek}{`o}
\DeclareMathSymbol{\spi}{\mathord}{sfgreek}{`p}
\DeclareMathSymbol{\srho}{\mathord}{sfgreek}{`r}
\DeclareMathSymbol{\ssigma}{\mathord}{sfgreek}{`s}
\DeclareMathSymbol{\stau}{\mathord}{sfgreek}{`t}
\DeclareMathSymbol{\supsilon}{\mathord}{sfgreek}{`u}
\DeclareMathSymbol{\sphi}{\mathord}{sfgreek}{`f}
\DeclareMathSymbol{\schi}{\mathord}{sfgreek}{`q}
\DeclareMathSymbol{\spsi}{\mathord}{sfgreek}{`y}
\DeclareMathSymbol{\somega}{\mathord}{sfgreek}{`w}

\DeclareMathSymbol{\svarsigma}{\mathord}{sfgreek}{`c}

\DeclareMathSymbol{\sGamma}{\mathalpha}{sfgreek}{`G}
\DeclareMathSymbol{\sDelta}{\mathalpha}{sfgreek}{`D}
\DeclareMathSymbol{\sTheta}{\mathalpha}{sfgreek}{`J}
\DeclareMathSymbol{\sLambda}{\mathalpha}{sfgreek}{`L}
\DeclareMathSymbol{\sXi}{\mathalpha}{sfgreek}{`X}
\DeclareMathSymbol{\sPi}{\mathalpha}{sfgreek}{`P}
\DeclareMathSymbol{\sSigma}{\mathalpha}{sfgreek}{`S}
\DeclareMathSymbol{\sUpsilon}{\mathalpha}{sfgreek}{`U}
\DeclareMathSymbol{\sPhi}{\mathalpha}{sfgreek}{`F}
\DeclareMathSymbol{\sPsi}{\mathalpha}{sfgreek}{`Y}
\DeclareMathSymbol{\sOmega}{\mathalpha}{sfgreek}{`W}

\DeclareRobustCommand{\mcal}[1]{%
  \ifcat\noexpand#1\relax\mathnormal{#1}\else\cal{#1}\fi
}
\DeclareRobustCommand{\BM}[1]{%
  \ifcat\noexpand#1\relax\bm{\boldUppercaseItalicGreek{#1}}\else\bm{#1}\fi
}
\makeatletter
\newcommand{\boldUppercaseItalicGreek}[1]{\csname var\expandafter\@gobble\string#1\endcsname}
\makeatother
\newcommand{\RS}[1]{\MakeUppercase{\msf{#1}}} 

\definecolor{BLUE}{rgb}{0,0,1}

\newtheorem{theorem}{Theorem}
\newtheorem{proposition}{Proposition}

\newtheorem{definition}{Definition}
\newtheorem{remark}{Remark}
\newtheorem{lemma}{Lemma}

\newtheorem{assumption}{Assumption}

\newcommand{\paperTitle}{A Minimax Framework for Two-Agent Scheduling with Inertial Constraints}

\acrodef{gnss}[GNSS]{global navigation satellite system}
\acrodef{rf}[RF]{radio frequency}
\acrodef{aoa}[AOA]{angle-of-arrival}
\acrodef{rss}[RSS]{received signal strength}
\acrodef{toa}[TOA]{time-of-arrival}
\acrodef{tdoa}[TDOA]{time-difference-of-arrival}
\acrodef{rtt}[RTT]{round-trip time}
\acrodef{fdd}[FDD]{frequency division duplex}
\acrodef{tdd}[TDD]{time division duplex}
\acrodef{fd}[FD]{full-duplex}
\acrodef{sdp}[SDP]{semidefinite programming}
\acrodef{crlb}[CRLB]{Cram\'{e}r-Rao lower bound}
\acrodef{nc}[NC]{narrow correlator}
\acrodef{sc}[SC]{storbe correlator}
\acrodef{pll}[PLL]{phase locked loop}
\acrodef{mp}[MP]{multipath}
\acrodef{sp}[SP]{single path}
\acrodef{ff}[FF]{flat fading}
\acrodef{mds}[MDS]{multidimensional scaling}
\acrodef{snr}[SNR]{signal-to-noise ratio}
\acrodef{los}[LOS]{line-of-sight}
\acrodef{nlos}[NLOS]{non-line-of-sight}
\acrodef{sic}[SIC]{serial interference cancelation}
\acrodef{pic}[PIC]{parallel interference cancelation}
\acrodef{adc}[ADC]{analog-to-digital converter}
\acrodef{bp}[BP]{basis pursuit}
\acrodef{lasso}[LASSO]{least absolute shrinkage and selection operator}
\acrodef{omp}[OMP]{orthogonal matching pursuit}
\acrodef{lls}[LLS]{linear least squares}
\acrodef{wlls}[WLLS]{weighted linear least squares}
\acrodef{nlls}[NLLS]{nonlinear least squares}
\acrodef{awgn}[AWGN]{additive white Gaussian noise}
\acrodef{cirf}[CIRF]{channel impulse response function}
\acrodef{irf}[IRF]{impulse response function}
\acrodef{llr}[LLR]{log-likelihood ratio}
\acrodef{llrs}[LLRs]{log-likelihood ratios}
\acrodef{fim}[FIM]{Fisher information matrix}
\acrodef{efim}[EFIM]{equivalent Fisher information matrix}
\acrodef{mse}[MSE]{mean squared error}
\acrodef{peb}[PEB]{position error bound}
\acrodef{rmse}[RMSE]{root mean squared error}
\acrodef{seb}[SEB]{synchronization error bound}
\acrodef{imu}[IMU]{inertial measurement unit}
\acrodef{nls}[NLS]{network localization and synchronization}
\acrodef{Nls}[NLS]{Network localization and synchronization}
\acrodef{reb}[REB]{ranging error bound}
\acrodef{co}[CO]{clock offset}
\acrodef{cdma}[CDMA]{code-division multiple-access}
\acrodef{pdf}[PDF]{probability density function}%

\acrodef{ppp}[PPP]{Poisson point process}%
\acrodef{gps}[GPS]{global positioning system}%
\acrodef{td}[$2$-D]{two-dimensional}%
\acrodef{thd}[$3$-D]{three-dimensional}%
\acrodef{dd}[$d$-D]{$d$-dimensional}
\acrodef{uav}[UAV]{unmanned aerial vehicle}%
\acrodef{iot}[IoT]{Internet of Things}%

\begin{document}

\bstctlcite{IEEEexample:BSTcontrol}
\title{\paperTitle}
\author{
	\vspace{0.2cm}
 Feihong~Yang,~\IEEEmembership{Graduate Student~Member,~IEEE}, and
 Yuan~Shen,~\IEEEmembership{Senior Member,~IEEE}

    \thanks{
        The authors are with the Department of Electronic Engineering, Tsinghua University, and Beijing National Research Center for Information Science and Technology, Beijing 100084, China (e-mail: {yfh17@mails.tsinghua.edu.cn}, {shenyuan\_ee@tsinghua.edu.cn}).
    }
}
\maketitle

\begin{abstract}
Autonomous agents are promising in applications such as intelligent transportation and smart manufacturing, and scheduling of agents has to take their inertial constraints into consideration. Most current researches require the obedience of all agents, which is hard to achieve in non-dedicated systems such as traffic intersections. In this article, we establish a minimax framework for the scheduling of two inertially constrained agents with no cooperation assumptions. Specifically, we first provide a unified and sufficient representation for various types of situation information, and define a state value function characterizing the agent's preference of states under a given situation. Then, the minimax control policy along with the calculation methods is proposed which optimizes the worst-case state value function at each step, and the safety guarantee of the policy is also presented. Furthermore, several generalizations are introduced on the applicable scenarios of the proposed framework. Numerical simulations show that the minimax control policy can reduce the largest scheduling cost by $13.4\%$ compared with queueing and following policies. Finally, the effects of decision period, observation period and inertial constraints are also numerically discussed.
\end{abstract}


\acresetall

\acresetall		

\section{Introduction}\label{secT:intro}
Advancing communication and control technologies enable the proliferation of autonomous agents such as robots, intelligent vehicles and unmanned aerial vehicles (UAVs). Compared with static devices, their dynamics greatly extend the scope of industrial and civil applications to scenarios including environment sensing, intelligent transportation and smart manufacturing \cite{WanTanShuLiWanImrVas:J16,SisSaiHanJenGid:J18,Olf:J06,KarAltEkiHeiJarLinWei:J11, LevAskBecDolHelKamKolLanPinPraSokStaStaTeiWerThr:C11,GupJaiVas:J16}. In multi-agent systems, abundant coordination and competition schemes can be developed for different tasks, which attract massive attention in various researches \cite{OlfFaxMur:J07,MarArsSha:J09,CaoYuRenChe:J13,CaiShe:J19}.

\begin{figure} [!t]
  \centering%
  \subfigure[]{%
    \centering
    \includegraphics[width=0.48\textwidth]{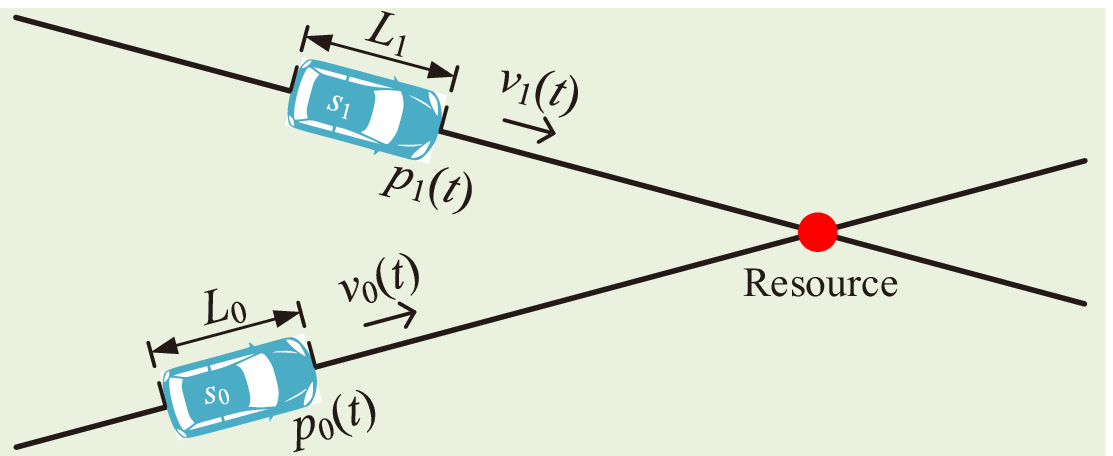}}\\
  \subfigure[]{%
    \centering
    \includegraphics[width=0.48\textwidth]{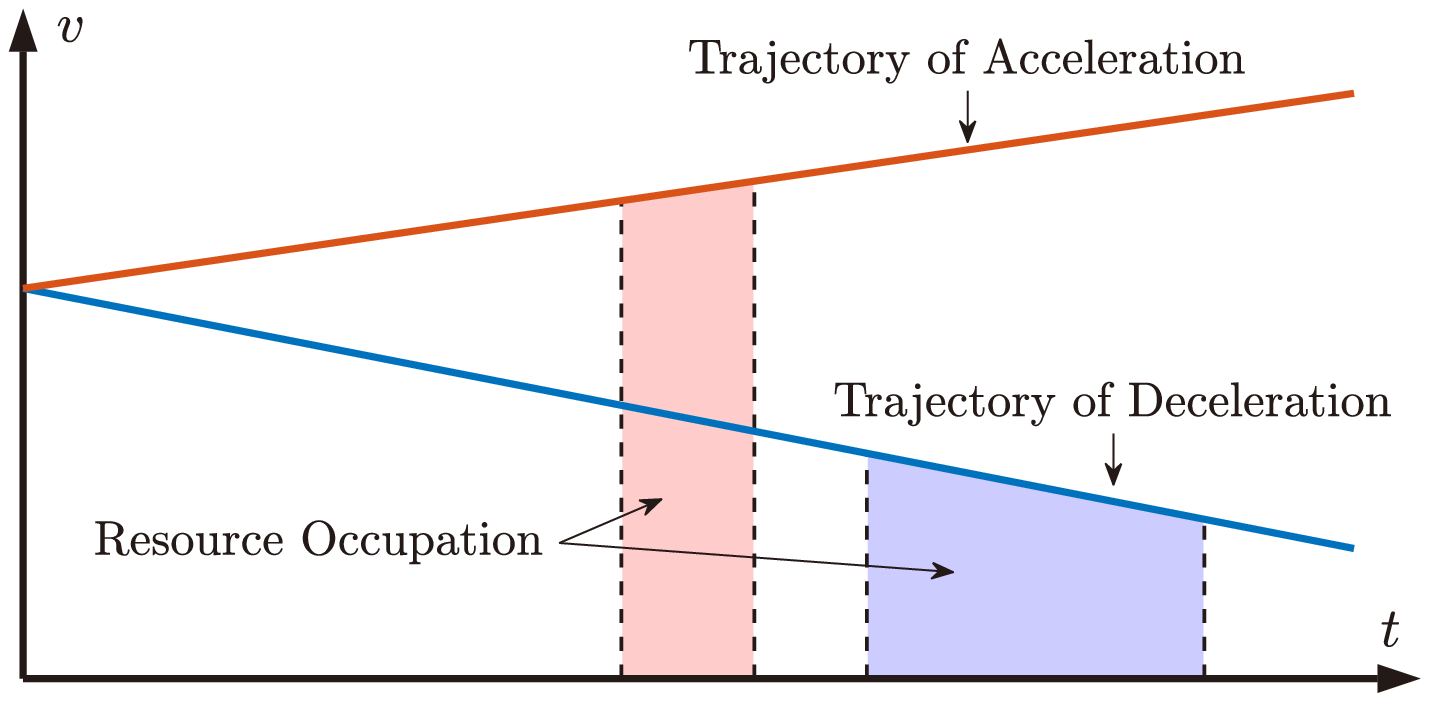}}
  \caption{(a) Two agents requiring a unique resource. (b) For an agent with inertial constraints, choosing different trajectories can result in different durations of the resource occupation. As a comparison, agents without inertial constraints can always reach the highest velocity in an instant, and the resource occupation durations can be regarded as a constant.}
  \label{figT:sys_1}
\end{figure}

When multiple agents in a system require a limited resource, scheduling schemes are necessary to avoid conflicts, as shown in Fig. \ref{figT:sys_1}(a). The research of scheduling problems has a long history \cite{ConMaxMil:B67,GraLawLenKan:J79,BruDreMohNeu:J99,MerMidSch:J02}, where the scheduling is traditionally modelled as optimization problems over the occupation time periods of the agents, and commonly-used metrics include (maximum or average) completion time and lateness. In recent studies of wireless communications, scheduling problems are also widely considered over data packets \cite{TasEph:J92,WuZenZha:J18,KadSinUysSinMod:J18,MahUmaSub:J19}, and some new metrics such as age of information, system capacity and energy consumption are proposed.

However, it is a different issue when the resource is space-related and the autonomous agents are constrained by inertia when approaching the resource, and one typical scenario is a traffic intersection in an intelligent transportation system (ITS). In this case, we cannot model agents by a set of time parameters such as the arrival time and the resource occupation duration as traditional scheduling problems, since these parameters are no longer predetermined and rely on the trajectories of agents, as shown in Fig. \ref{figT:sys_1}(b). Instead, the spatial states of the agents need to be characterized in the modelling including their positions and velocities, and detailed agent trajectories need to be generated in a scheduling policy besides the time slot assignments of the resource. Furthermore, since the velocities of agents during the occupation is not constant, they should also be considered when evaluating the scheduling performance.


There are some works taking the inertial constraints into consideration, a large portion of which lie under the scenario of scheduling intelligent vehicles near traffic intersections \cite{CheEng:J16,RioMal:J17}. Generally, these researches can be divided into two main categories. The first category applies a hybrid framework, where the occupation time periods (possibly with redundancy) or an occupation order is firstly assigned to the agents by a central device, and then the control policies are designed for single agents to meet the requirements. Specifically, protocols such as adaptive traffic lights \cite{ZhaDaiZha:J12,GreQiaFraForWon:J15,YanRakAla:J17,XuBanBiaLiWanLiLi:J19,PhaNgoLe:J20}, time slot reservation \cite{KowCavKum:J11,HuaSadZha:J12,ZhaZhaForWuGre:J15,YanShe:C20} and central optimization \cite{WuPerAbb:J14,LinLiuJinRan:J17,HulZanGroWymFal:J20,TalCor:J20,MedCreLefWou:J20} are analyzed to complete the first step of scheduling. Furthermore, there are also some researches focusing on the distributed implementations of these protocols \cite{DreSto:J08,AhmAbbPerWuMouBuiZeo:J13,WuZhaLuoCao:J15}, in which the central device can be removed. Another category contains the trajectory planning frameworks, which directly return the non-conflicting trajectories of all agents without intermediate steps using methods including characterizing the maximal controlled invariant set \cite{HafDel:J11,ColDel:J15} and performing optimization over the control inputs \cite{GlaVanMamGruNou:J10,LeePar:J12,CamFalSjo:C13,KamImuHayOhaAih:J15,ChoNazWymCha:C19}.


However, most of current researches are based on the assumption that all agents in the system follow the proposed protocol, while loading the protocol to all agents often costs much in a complex system, and is sometimes impossible in non-dedicated systems such as traffic intersections containing human-driven vehicles. As long as a disobedient agent exists, the safety of the system cannot be guaranteed. Exceptions such as \cite{DreSto:J08,HafDel:J11,LinLiuJinRan:J17} allow the existence of non-autonomous agents, while they only focus on finding a collision avoiding solution without considering the scheduling performance.


In this article, we consider the scheduling problem of inertially constrained agents from the one-agent perspective, to which the other agents are not necessarily cooperative. Specifically, we adopt the two-agent model for simplicity, which is sufficient to illustrate the main idea and obtain non-trivial insights. The contributions of this article are as follows.
\begin{itemize}
\item We propose a scheduling framework taking the acceleration limits of agents into account and develop a unified and sufficient representation of various types of situation information for the scheduling of an agent.
\item We establish a state value function for an agent based on its obtained situation information by minimizing the manageable component of the accessible cost, which characterizes its preference of different states for the scheduling.
\item We develop a minimax control policy for the scheduling problem which accords with the intuition of moving toward the state with the best worst-case preference, and further prove the safety guarantee of the policy and verify its robust performance by comparing with other typical policies through numerical simulations.
\end{itemize}

The rest of the article is organized as follows. Section \ref{secT:model} formulates the two-agent scheduling problem by introducing the scheduling cost. Then in Section \ref{secT:theo}, we characterize the unified representation of situation information and propose the state value function under a fixed situation. The minimax control policy is established in Section \ref{secT:policy} along with the safety guarantee and the calculation methods. Section \ref{secT:disc} provides intuitions and several generalizations of the framework, and Section \ref{secT:result} presents numerical results on the state value function and the scheduling performance of the proposed policy. Finally, the article is concluded in Section \ref{secT:conclu}.

\emph{Notations:} Throughout the article, when we want to refer to a variable, such as $p_i(t)$ (the position of agent $i$ at time $t$), with emphasizing the corresponding trajectory which, for example, is denoted by $\sigma$, we use the notations with superscript $\sigma$, such as $p_i^{\sigma}(t)$. We use $(a,b)$ and $[a,b]$ to represent an open interval and a closed interval, respectively, and $\langle a,b\rangle$ is used to represent an ordered pair to avoid ambiguity. For a set $\mathcal{A}$, $\mathrm{card}(\mathcal{A})$ represents its cardinality and $\mu(\mathcal{A})$ represents its Lebesgue measure. For equations, ``$\triangleq$'' is used when one side is undefined and it represents the definition of this side.



\section{Modelling and Problem Formulation}\label{secT:model}
In this section, we introduce the system model and formulate the scheduling problem of two inertially constrained agents.

\subsection{System Model}\label{subsecT:model}
Consider a system with two agents $s_0$ and $s_1$, both of which move along a given track and require a unique resource in some future stages, as illustrated in Fig. \ref{figT:sys_1}(a).\footnote{The intersection shown in Fig. \ref{figT:sys_1}(a) is a topological intersection instead of a geometric one. In other words, issues such as the intersection angle of the tracks or the widths of the agents are beyond the scope of the article.} For each agent $s_i$ ($i\in\{0,1\}$), an axis is attached to characterize its movement, where the origin is always set to be the position of the resource. Let $L_i$ be the resource demand of the agent $s_i$ under its own axis, and let $p_i(t)$, $v_i(t)$ and $a_i(t)$ be the position, velocity and acceleration of $s_i$ at time $t$, respectively. Therefore, the resource is occupied by $s_i$ at time $t$ if and only if $p_i(t)\in(0,L_i)$.

Furthermore, we impose the following two kinematic constraints on the two agents that for $i\in\{0,1\}$ and $t\in\mathbb{R}_+$,
\begin{align}
0&\le v_i(t)\le v_{\mathrm{M}}\label{eqT:v_cons}\\
-a_{i,\mathrm{m}}&\le a_i(t)\le a_{i,\mathrm{M}}\label{eqT:a_cons}
\end{align}
where $v_{\mathrm{M}}$, $a_{i,\mathrm{m}}$, $a_{i,\mathrm{M}}$ are positive numbers.\footnote{Without loss of generality, the maximum velocity constraint $v_{\mathrm{M}}$ is set to be equal for the two agents, which can be achieved by rescaling the axis of either agent.} Specifically, note that \eqref{eqT:a_cons} is the \emph{inertial constraint} which is common among physical agents. Under the two constraints \eqref{eqT:v_cons} and \eqref{eqT:a_cons}, we can regard the pair
\begin{equation}
\mathbf{x}_i(t)\triangleq\langle p_i(t),v_i(t)\rangle
\end{equation}
as the \emph{state} and regard $a_i(t)$ as the \emph{control} of $s_i$ at time $t$.


To ensure the safety of the system, we do not allow the two agents to simultaneously occupy the resource, i.e.,
\begin{equation}\label{eqT:safety1}
\mathrm{card}\big\{i\,\big|\,p_i(t)\in(0,L_i)\big\}\le 1,\quad \forall\, t\ge0.
\end{equation}
Furthermore, by defining
\begin{align}
t_{i,\mathrm{in}}&\triangleq\inf\{t\,|\,p_i(t)>0\}\label{eqT:t_in}\\
t_{i,\mathrm{out}}&\triangleq\sup\{t\,|\,p_i(t)<L_i\}\label{eqT:t_out}
\end{align}
for $i\in\{0,1\}$, the safety condition is equivalent to
\begin{equation}\label{eqT:safety2}
(t_{0,\mathrm{in}},t_{0,\mathrm{out}})\cap(t_{1,\mathrm{in}},t_{1,\mathrm{out}})=\varnothing.
\end{equation}
Therefore, scheduling is required in order to meet the safety condition.

\subsection{Formulation of the Scheduling Problem}\label{subsecT:formu}
In this article, we consider the scheduling problem from the perspective of the agent $s_0$. In other words, we can only control $s_0$ and know nothing about the future control of $s_1$. Nevertheless, we assume that the constraints \eqref{eqT:v_cons} and \eqref{eqT:a_cons} are known by $s_0$. Furthermore, $s_0$ is allowed to obtain information on history or current states of $s_1$ by observation or by communication with other possible cooperative devices. However, due to the non-ideal environment, observation or communication may suffer from failure and delay, and thus $s_0$ cannot predict when and which information can be successfully obtained in the future.

To ensure the existence of a safe scheduling, we assume that $s_0$ is able to stop before the resource. In other words, the initial state of $s_0$ satisfies
\begin{equation}\label{eqT:initial}
p_0(0)+\frac{v_0(0)^2}{2a_{0,\mathrm{m}}}\le 0.
\end{equation}
We aim at developing a causal control policy for $s_0$ which generates a trajectory with a low scheduling cost. Specifically, we consider a discrete set of decision times $\{t_k\,|\,k\ge0\}$ with $t_0=0$ and $t_k<t_{k+1}$ for any $k\ge0$. Then at each decision time $t_k$, the policy can make use of all collected causal information and return the trajectory of $s_0$ on the next time period $[t_k,t_{k+1}]$. Now we define the scheduling cost of $s_0$ under a fixed trajectory of $s_1$.

\begin{definition}[Scheduling Cost]\label{defT:postcost}
Fix the trajectory of $s_1$. Then for a given trajectory $\sigma_0$ of $s_0$, the scheduling cost is defined by
\begin{align}
&C(\sigma_0;t_{1,\mathrm{in}},t_{1,\mathrm{out}})\notag\\
&\quad\triangleq
\begin{cases}v_{\mathrm{M}}t_{0,\mathrm{in}}^{\sigma_0}+ \frac{1}{2a_{0,\mathrm{M}}}\big(v_{\mathrm{M}}-v_0^{\sigma_0}(t_{0,\mathrm{in}}^{\sigma_0})\big)^2,&\\
&\hspace{-4cm}\text{if } (t_{0,\mathrm{in}}^{\sigma_0},t_{0,\mathrm{out}}^{\sigma_0})\cap(t_{1,\mathrm{in}},t_{1,\mathrm{out}})=\varnothing;\\
+\infty,&\hspace{-4cm}\text{otherwise}.
\end{cases}\label{eqT:postcost}
\end{align}
\end{definition}

\begin{figure} [!t]
  \centering%
  \includegraphics[width=0.48\textwidth]{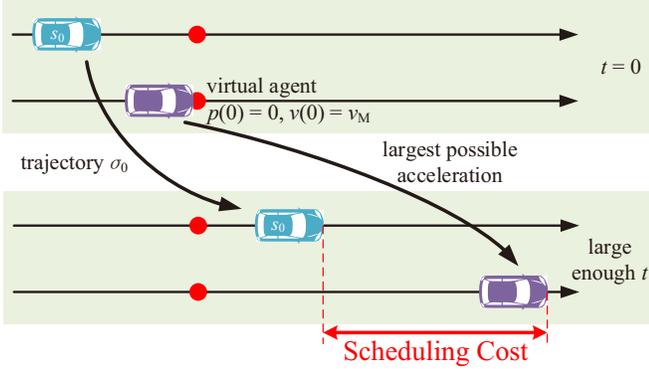}
  \caption{Intuition of the scheduling cost $C(\sigma_0;t_{1,\mathrm{in}},t_{1,\mathrm{out}})$.}
  \label{figT:sys_4}
\end{figure}

According to \eqref{eqT:postcost}, only the control before $t_{0,\mathrm{in}}$ makes a difference to the cost as long as the safety condition is satisfied. For completeness, we always set the control of $s_0$ to be
\begin{equation}
a_0(t)=a_{0,\mathrm{A}}(t)\triangleq\begin{cases}
a_{0,\mathrm{M}},&\text{if }v_0(t)<v_{\mathrm{M}}; \\
0,&\text{if }v_0(t)=v_{\mathrm{M}}\end{cases}
\end{equation}
for $t\ge t_{0,\mathrm{in}}$, which is always a good choice to reduce $t_{0,\mathrm{out}}$ and decrease the possibility of conflict.\footnote{Note that this is only the restriction of the trajectory of $s_0$. We cannot make the similar assumption for $s_1$ since we have no control of it.}

Furthermore, note that when the safety condition holds, the cost is in unit of distance and can be divided into two parts, which can be regarded as the ``time cost'' related to $t_{0,\mathrm{in}}$ and the ``velocity cost'' related to $v_0(t_{0,\mathrm{in}})$, respectively. This definition is intuitive since a earlier occupation time and a quicker occupation velocity implies a higher scheduling efficiency. It can also be interpreted as the final distance between $s_0$ with a virtual agent which starts from the state $p(0)=0$, $v(0)=v_{\mathrm{M}}$ and always follows the largest possible acceleration, as shown in Fig. \ref{figT:sys_4}.

In summary, the scheduling problem we consider in this article can be formulated as follows.

\vspace{0.2cm}
\textbf{Problem:} Designing a causal control policy for the agent $s_0$ to reduce the scheduling cost $C(\sigma_0;t_{1,\mathrm{in}},t_{1,\mathrm{out}})$, where $\sigma_0$ is the output trajectory of the policy.
\vspace{0.2cm}

\subsection{Sketch of the Article}\label{subsecT:ske}
Due to the causality constraint, $t_{1,\mathrm{in}}$ and $t_{1,\mathrm{out}}$ are unknown in advance, and thus the scheduling cost $C(\sigma_0;t_{1,\mathrm{in}},t_{1,\mathrm{out}})$ cannot be directly accessed by $s_0$. To tackle the difficulty, we propose a control policy (Algorithm \ref{algT:1}) in Section \ref{subsecT:statement}, in which the most crucial step is solving a minimax optimization problem \eqref{eqT:opt}. Intuitively, under the proposed policy, $s_0$ always moves toward the state with higher preference based on its currently obtained situation information.

\begin{figure} [!t]
  \centering%
  \includegraphics[width=0.48\textwidth]{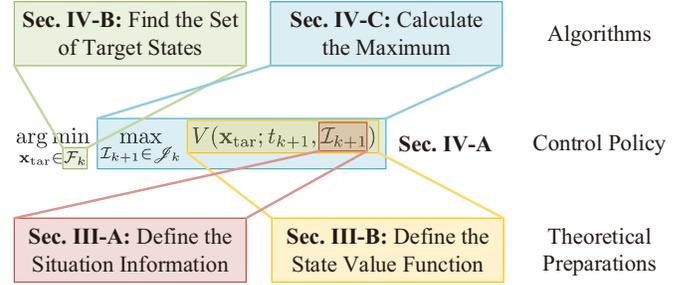}
  \caption{Sketch of the article.}
  \label{figT:sys_6}
\end{figure}

The sketch of the article is shown in Fig. \ref{figT:sys_6}. We first formulate and analyze the concepts of ``situation information'' and ``preference of states'' in Section \ref{secT:theo} as theoretical preparations. Then we establish the minimax control policy and provide the safety guarantee in Section \ref{subsecT:statement}. After that, algorithms for the implementation of the policy are developed.

\section{Theoretical Preparations}\label{secT:theo}
In this section, we characterize the obtained situation information from various sources with a unified form denoted by $\mathcal{I}$, and then establish the state value function for a given situation.

Recall that all trajectories considered in this article need to satisfy the kinematic constraints \eqref{eqT:v_cons} and \eqref{eqT:a_cons}. For convenience, we first provide some definitions on the agent trajectories.

\begin{definition}[Trajectory Set]
Let $t_0\ge0$ and $\mathbf{x}=\langle p,v\rangle$ with $0\le v\le v_{\mathrm{M}}$. Then for an agent $s_i$, define
\begin{equation}
\Sigma_i(\mathbf{x};t_0)\triangleq\left\{\sigma\,\middle| \,\begin{aligned}
\mathbf{x}_i^{\sigma}(t_0)&=\mathbf{x}\\
0\le v_i^{\sigma}(t)&\le v_{\mathrm{M}},\forall\, t\ge0\\
-a_{i,\mathrm{m}}\le a_i^{\sigma}(t)&\le a_{i,\mathrm{M}},\forall\, t\ge0
\end{aligned}\right\}
\end{equation}
be the set of all trajectories of $s_i$ with the state $\mathbf{x}$ at time $t_0$.
\end{definition}

\begin{definition}[Typical Trajectories]\label{defT:traj}
Let $\mathcal{T}\subseteq\mathbb{R}_+$ be an interval and let
\begin{align}
a_{i,\mathrm{A}}(t)&\triangleq\begin{cases}a_{i,\mathrm{M}},&\text{if }v_i(t)<v_{\mathrm{M}}; \\0,&\text{if }v_i(t)=v_{\mathrm{M}}\end{cases}\label{eqT:aA}\\
a_{i,\mathrm{D}}(t)&\triangleq\begin{cases}-a_{i,\mathrm{m}},&\text{if }v_i(t)>0; \\0,&\text{if }v_i(t)=0.\end{cases}\label{eqT:aD}
\end{align}
Then a trajectory $\sigma$ of $s_i$ is called
\begin{itemize}
\item a \textsc{Dec} trajectory on $\mathcal{T}$, if $a_i^{\sigma}(t)=a_{i,\mathrm{D}}(t)$ for all $t\in\mathcal{T}$;
\item an \textsc{Acc} trajectory on $\mathcal{T}$, if $a_i^{\sigma}(t)=a_{i,\mathrm{A}}(t)$ for all $t\in\mathcal{T}$;
\item a \textsc{Dec}-\textsc{Acc} trajectory on $\mathcal{T}$, if there exists $t_{\mathrm{D}}^{\sigma}\in\mathcal{T}$, such that $a_i^{\sigma}(t)=a_{i,\mathrm{D}}(t)$ for $t\in\mathcal{T}\cap(0,t_{\mathrm{D}}^{\sigma})$ and $a_i^{\sigma}(t)=a_{i,\mathrm{A}}(t)$ for $t\in\mathcal{T}\cap(t_{\mathrm{D}}^{\sigma},+\infty)$;
\item an \textsc{Acc}-\textsc{Dec} trajectory on $\mathcal{T}$, if there exists $t_{\mathrm{A}}^{\sigma}\in\mathcal{T}$, such that $a_i^{\sigma}(t)=a_{i,\mathrm{A}}(t)$ for $t\in\mathcal{T}\cap(0,t_{\mathrm{A}}^{\sigma})$ and $a_i^{\sigma}(t)=a_{i,\mathrm{D}}(t)$ for $t\in\mathcal{T}\cap(t_{\mathrm{A}}^{\sigma},+\infty)$.
\end{itemize}
\end{definition}

\subsection{Information Extraction}\label{subsecT:extrac}
In this subsection, we fix the current time $t_{\mathrm{cur}}$ and aim at extracting useful information for the scheduling task of $s_0$. We assume that $p_0(t_{\mathrm{cur}})\le0$; otherwise, the scheduling is unnecessary. Furthermore, the following technical assumption is made to avoid the infinite occupation time of $s_1$.
\begin{assumption}\label{asT:1}
$t_{1,\mathrm{out}}\le B$, where $B$ is a large enough positive number.
\end{assumption}

Note that $p_0(t_{\mathrm{cur}})\le0$ implies $t_{\mathrm{cur}}\le t_{0,\mathrm{in}}\le t_{0,\mathrm{out}}$. According to the scheduling cost defined in \eqref{eqT:postcost}, it is easy to check that
\begin{equation}\label{eqT:C=C}
C(\sigma_0;t_{1,\mathrm{in}},t_{1,\mathrm{out}})=C(\sigma_0;\hat{t}_{1,\mathrm{in}},\hat{t}_{1,\mathrm{out}})
\end{equation}
where
\begin{equation}\label{eqT:hat}
\hat{t}_{1,\mathrm{in}}\triangleq \max\{t_{1,\mathrm{in}},t_{\mathrm{cur}}\},\quad
\hat{t}_{1,\mathrm{out}}\triangleq \max\{t_{1,\mathrm{out}},t_{\mathrm{cur}}\}.
\end{equation}
In other words, all information that $s_0$ needs for its scheduling at time $t_{\mathrm{cur}}$ is contained in the pair $\langle\hat{t}_{1,\mathrm{in}},\hat{t}_{1,\mathrm{out}}\rangle$.


Although the exact value of $\langle\hat{t}_{1,\mathrm{in}},\hat{t}_{1,\mathrm{out}}\rangle$ is generally unknown at time $t_{\mathrm{cur}}$, a subset $\mathcal{I}\subseteq\mathbb{R}^2$ in which $\langle\hat{t}_{1,\mathrm{in}},\hat{t}_{1,\mathrm{out}}\rangle$ must lie can always be obtained. Therefore, we can use this uncertainty set to characterize the \emph{situation information} obtained by $s_0$ at $t_{\mathrm{cur}}$. Specifically, if nothing is known about $s_1$ at $t_{\mathrm{cur}}$, we can let
\begin{equation}\label{eqT:basicinfo}
\mathcal{I}=
\left\{\langle\hat{t}_{1,\mathrm{in}},\hat{t}_{1,\mathrm{out}}\rangle\,\middle|\,\begin{aligned}
&\hat{t}_{1,\mathrm{in}}=\max\{t_{1,\mathrm{in}},t_{\mathrm{cur}}\}\\
&\hat{t}_{1,\mathrm{out}}=\max\{t_{1,\mathrm{out}},t_{\mathrm{cur}}\}\\
&t_{1,\mathrm{out}}\le B\\
&0\le t_{1,\mathrm{in}}\le t_{1,\mathrm{out}}-L_1/v_{\mathrm{M}}
\end{aligned}\right\}.
\end{equation}
When some information on $s_1$ is available, $\mathcal{I}$ can be narrowed to a subset of \eqref{eqT:basicinfo}, and different types of information can be fused by intersecting their respective subsets.

In the next two propositions, we specifically calculate and characterize the uncertainty set $\mathcal{I}$ for the scenario where the \emph{accurate state} of $s_1$ at time $t_{\mathrm{obs}}\le t_{\mathrm{cur}}$ is observed. Note that this scenario allows a positive age of information and is common in various practical scenarios.

\begin{figure} [!t]
  \centering%
  \includegraphics[width=0.48\textwidth]{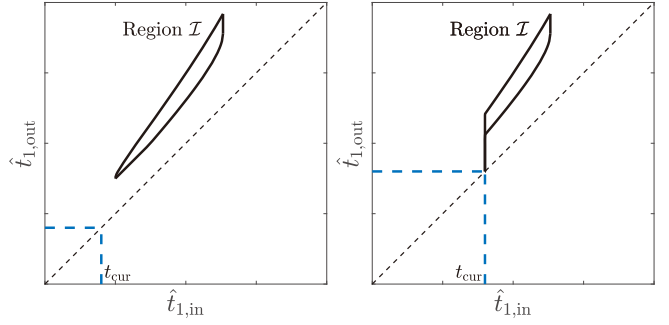}
  \caption{The uncertainty set $\mathcal{I}$ calculated by Proposition \ref{propT:1} at different $t_{\mathrm{cur}}$ based on the same observation.}
  \label{figT:theo_1_1}
\end{figure}

\begin{proposition}\label{propT:1}
Suppose $\mathbf{x}_1(t_{\mathrm{obs}})=\mathbf{x}_*$ is observed at time $t_{\mathrm{cur}}$, where $t_{\mathrm{obs}}\le t_{\mathrm{cur}}$. Then for any trajectory of $s_1$ according with the observation, $\langle\hat{t}_{1,\mathrm{in}},\hat{t}_{1,\mathrm{out}}\rangle$ must lie in $\mathcal{I}$, where $\mathcal{I}$ is defined as follows.
\begin{itemize}
\item[(i)] If $p_1(t_{\mathrm{obs}})\le 0$, then
    \begin{equation}\label{eqT:prop1_1}
    \mathcal{I}=\left\{\langle\hat{t}_{1,\mathrm{in}},\hat{t}_{1,\mathrm{out}}\rangle\,\middle| \,\begin{aligned}
    &\text{Conditions in }\eqref{eqT:basicinfo}\\
    &t_{1,\mathrm{in}}^{\sigma_{\mathrm{A}}}\le t_{1,\mathrm{in}}\le t_{1,\mathrm{in}}^{\sigma_{\mathrm{D}}}\\
    &t_{1,\mathrm{out}}^{\sigma_{\mathrm{DA}}(t_{1,\mathrm{in}})}\le t_{1,\mathrm{out}}\le t_{1,\mathrm{out}}^{\sigma_{\mathrm{AD}}(t_{1,\mathrm{in}})}
    \end{aligned}\right\}
    \end{equation}
    where $\sigma_{\mathrm{A}},\sigma_{\mathrm{D}}\in\Sigma_1(\mathbf{x}_*;t_{\mathrm{obs}})$ are the \textsc{Acc} trajectory and the \textsc{Dec} trajectory on $[t_{\mathrm{obs}},+\infty)$, respectively; for any $t_*$,
    $\sigma_{\mathrm{DA}}(t_*),\sigma_{\mathrm{AD}}(t_*)\in\Sigma_1(\mathbf{x}_*;t_{\mathrm{obs}})$ are the \textsc{Dec}-\textsc{Acc} trajectory and the \textsc{Acc}-\textsc{Dec} trajectory on $[t_{\mathrm{obs}},+\infty)$ satisfying $t_{1,\mathrm{in}}^{\sigma_{\mathrm{DA}}(t_*)}= t_{1,\mathrm{in}}^{\sigma_{\mathrm{AD}}(t_*)}=t_*$, respectively.\footnote{For the last inequality in \eqref{eqT:prop1_1}, if the trajectory $\sigma_{\mathrm{DA}}(t_{1,\mathrm{in}})$ or $\sigma_{\mathrm{AD}}(t_{1,\mathrm{in}})$ does not exist for some $t_{1,\mathrm{in}}$, we omit the corresponding side of the inequality.}
\item[(ii)] If $0<p_1(t_{\mathrm{obs}})<L_1$, then
    \begin{equation}
    \mathcal{I}=\left\{\langle\hat{t}_{1,\mathrm{in}},\hat{t}_{1,\mathrm{out}}\rangle\,\middle|\,\begin{aligned}
    &\text{Conditions in }\eqref{eqT:basicinfo}\\
    &t_{1,\mathrm{in}}\le t_{\mathrm{cur}}\\
    &t_{1,\mathrm{out}}^{\sigma_{\mathrm{A}}}\le t_{1,\mathrm{out}}\le t_{1,\mathrm{out}}^{\sigma_{\mathrm{D}}}
    \end{aligned}\right\}
    \end{equation}
    where $\sigma_{\mathrm{A}},\sigma_{\mathrm{D}}\in\Sigma_1(\mathbf{x}_*;t_{\mathrm{obs}})$ are the \textsc{Acc} trajectory and the \textsc{Dec} trajectory on $[t_{\mathrm{obs}},+\infty)$, respectively.
\item[(iii)] If $p_1(t_{\mathrm{obs}})\ge L_1$, then $\mathcal{I}=\{\langle t_{\mathrm{cur}},t_{\mathrm{cur}}\rangle\}$.
\end{itemize}
\end{proposition}

\begin{IEEEproof}
See Appendix \ref{subsecT:pf_prop1}.
\end{IEEEproof}

\begin{proposition}\label{propT:1a}
Let $\mathcal{I}$ be the set defined in Proposition \ref{propT:1}. Then $\mathcal{I}$ is closed and simply connected, and $\mathcal{I}\cap\{\hat{t}_{1,\mathrm{in}}=t_*\}$ is either empty or a closed interval for any $t_*$. Furthermore, if $\mathcal{I}\cap\{\hat{t}_{1,\mathrm{in}}=t_*\}$ is non-empty, let
\begin{align}
\underline{\hat{t}_{1,\mathrm{out}}}(t_*)&\triangleq\min\big(\mathcal{I}\cap\{\hat{t}_{1,\mathrm{in}}=t_*\}\big)\\
\overline{\hat{t}_{1,\mathrm{out}}}(t_*)&\triangleq\max\big(\mathcal{I}\cap\{\hat{t}_{1,\mathrm{in}}=t_*\}\big).
\end{align}
Then both $\underline{\hat{t}_{1,\mathrm{out}}}(t_*)-t_*$ and $\overline{\hat{t}_{1,\mathrm{out}}}(t_*)-t_*$ are positive and non-decreasing functions with respect to $t_*$.
\end{proposition}

\begin{IEEEproof}
See Appendix \ref{subsecT:pf_prop1a}.
\end{IEEEproof}

Fig. \ref{figT:theo_1_1} provides two examples of the uncertainty set $\mathcal{I}$ calculated by Proposition \ref{propT:1}, from which the properties in Proposition \ref{propT:1a} can be intuitively checked. The last result in Proposition \ref{propT:1a} implies that $\hat{t}_{1,\mathrm{out}}-\hat{t}_{1,\mathrm{in}}$ is generally positively correlated with $\hat{t}_{1,\mathrm{in}}$. This result accords with intuition since a smaller $\hat{t}_{1,\mathrm{in}}$ often corresponds to a trajectory with more acceleration, which can also lead to a smaller $\hat{t}_{1,\mathrm{out}}-\hat{t}_{1,\mathrm{in}}$.

The results in Proposition \ref{propT:1a} can provide much convenience for the further analyses, while it is based on the scenario of accurate state observation. Therefore, we make the following assumption for the general case.
\begin{assumption}\label{asT:2}
$\mathcal{I}\subseteq\mathbb{R}^2$ always satisfies Proposition \ref{propT:1a}.
\end{assumption}

\begin{figure} [!t]
  \centering%
  \includegraphics[width=0.48\textwidth]{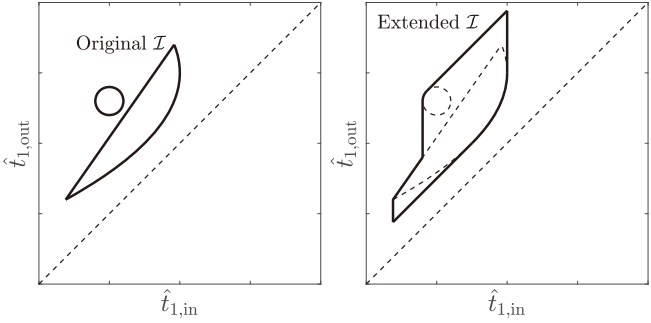}
  \caption{An extension of a general uncertainty set $\mathcal{I}$ which makes it satisfy Assumption \ref{asT:2}.}
  \label{figT:theo_1_2}
\end{figure}

It is easy to check that besides the case of accurate state information, the maximum uncertainty set given in \eqref{eqT:basicinfo} also satisfies Assumption \ref{asT:2}. For a general uncertainty set to which Assumption \ref{asT:2} may not hold, set extensions can be performed to make it satisfy the assumption. An example of the extension is illustrated in Fig. \ref{figT:theo_1_2}.

\subsection{State Value Function}\label{subsecT:obj}
In this subsection, we still consider a fixed time $t_{\mathrm{cur}}$ with $p_0(t_{\mathrm{cur}})\le0$, and propose a state value function for $s_0$ under a given situation information $\mathcal{I}$.

We start from the scheduling cost defined in \eqref{eqT:postcost}. We have shown in \eqref{eqT:C=C} that $\langle \hat{t}_{1,\mathrm{in}},\hat{t}_{1,\mathrm{out}}\rangle$ can be used to replace $\langle t_{1,\mathrm{in}},t_{1,\mathrm{out}}\rangle$. However, note that $\langle \hat{t}_{1,\mathrm{in}},\hat{t}_{1,\mathrm{out}}\rangle$ is unknown, and based on the obtained situation information, $s_0$ can only access the following function
\begin{align}
\!\!\!C_{\mathrm{max}}(\sigma_0;\mathcal{I}) &\triangleq\max_{\langle\hat{t}_{1,\mathrm{in}},\hat{t}_{1,\mathrm{out}}\rangle\in\mathcal{I}}\; C(\sigma_0;\hat{t}_{1,\mathrm{in}},\hat{t}_{1,\mathrm{out}})\notag\\
&=\begin{cases}v_{\mathrm{M}}t_{0,\mathrm{in}}^{\sigma_0}+ \frac{1}{2a_{0,\mathrm{M}}}\big(v_{\mathrm{M}}-v_0^{\sigma_0}(t_{0,\mathrm{in}}^{\sigma_0})\big)^2,&\\
&\hspace{-4.4cm}\text{if } (t_{0,\mathrm{in}}^{\sigma_0},t_{0,\mathrm{out}}^{\sigma_0})\cap\mathcal{W}(\mathcal{I})=\varnothing;\\
+\infty,&\hspace{-4.4cm}\text{otherwise}\label{eqT:accesscost}
\end{cases}\!\!\!
\end{align}
where
\begin{equation}
\mathcal{W}(\mathcal{I})\triangleq \bigcup_{\langle\hat{t}_{1,\mathrm{in}},\hat{t}_{1,\mathrm{out}}\rangle\in\mathcal{I}} (\hat{t}_{1,\mathrm{in}},\hat{t}_{1,\mathrm{out}}).\label{eqT:W}
\end{equation}
Specifically, we refer to the condition
\begin{equation}\label{eqT:safety3}
(t_{0,\mathrm{in}},t_{0,\mathrm{out}})\cap\mathcal{W}(\mathcal{I})=\varnothing
\end{equation}
as the \emph{robust safety condition} under the situation information $\mathcal{I}$ to make a distinction with the safety condition \eqref{eqT:safety2}.

A further observation on \eqref{eqT:accesscost} is that the time-related part $v_{\mathrm{M}}t_{0,\mathrm{in}}^{\sigma}$ contains two constant and ineliminable components for a fixed $t_{\mathrm{cur}}$ and $\mathcal{I}$. Specifically, $s_0$ is not able to occupy the resource in $[0,t_{\mathrm{cur}}]$ due to causality, and is also not able to start the occupation during $\mathcal{W}(\mathcal{I})$ due to the robust safety condition. Therefore, for the purpose of performing state evaluation, we drop these uncontrollable components and define the \emph{manageable cost} as follows.

\begin{definition}[Manageable Cost]\label{defT:priorcost}
Let $\mathcal{I}$ be the obtained situation information at time $t_{\mathrm{cur}}$. Then for a given trajectory $\sigma_0$ of $s_0$, the manageable cost is defined by
\begin{align}\label{eqT:priorcost}
&C^*(\sigma_0;t_{\mathrm{cur}},\mathcal{I})\notag\\
&\triangleq \begin{cases}
v_{\mathrm{M}}\delta(\sigma_0;t_{\mathrm{cur}},\mathcal{I}) +\frac{1}{2a_{0,\mathrm{M}}}\big(v_{\mathrm{M}}-v_0^{\sigma_0}(t_{0,\mathrm{in}}^{\sigma_0})\big)^2,&\\
&\hspace{-4.6cm}\text{if }(t_{0,\mathrm{in}}^{\sigma_0},t_{0,\mathrm{out}}^{\sigma_0})\cap\mathcal{W}(\mathcal{I})=\varnothing;\\
+\infty,&\hspace{-4.6cm}\text{otherwise}\end{cases}
\end{align}
where
\begin{equation}
\delta(\sigma_0;t_{\mathrm{cur}},\mathcal{I})\triangleq \mu\big[(t_{\mathrm{cur}},t_{0,\mathrm{in}}^{\sigma_0})\setminus\mathcal{W}(\mathcal{I})\big].
\end{equation}
\end{definition}


To characterize the preference of different states, we introduce the concept of \emph{state value function}, whose name is borrowed from the reinforcement learning researches. Specifically, for a given state, its value is defined by minimizing the manageable cost over all trajectories of $s_0$ starting from this state, as shown in the next definition.

\begin{definition}[State Value Function]\label{defT:evalfunc}
Let $\mathcal{I}$ be the obtained situation information at time $t_{\mathrm{cur}}$. Then the state value function of $s_0$ is defined by
\begin{equation}\label{eqT:evalfunc}
V(\mathbf{x};t_{\mathrm{cur}},\mathcal{I}) \triangleq\min_{\sigma_0\in\Sigma_0(\mathbf{x};t_{\mathrm{cur}})}C^*(\sigma_0;t_{\mathrm{cur}},\mathcal{I}).
\end{equation}
\end{definition}

Intuitively, states with smaller $V$ are more appealing based on the situation information $\mathcal{I}$, since they have a potential to result in a smaller manageable cost $C^*$. However, \eqref{eqT:evalfunc} is defined by a minimization problem and is indirect to calculate. Thus, we provide the next proposition characterizing a solution of the minimization in \eqref{eqT:evalfunc}. Specifically, the problem of calculating the state value function is transformed to calculating the manageable cost for a specific \textsc{Dec}-\textsc{Acc} trajectory.

\begin{proposition}\label{propT:2}
Let $\Lambda$ be the set of \textsc{Dec}-\textsc{Acc} trajectories on $[t_{\mathrm{cur}},+\infty)$ in $\Sigma_0(\mathbf{x};t_{\mathrm{cur}})$ satisfying the robust safety condition \eqref{eqT:safety3}. If $\Lambda\neq\varnothing$, let $\sigma^*\triangleq\mathop{\arg\min}_{\sigma\in\Lambda}\;t_{\mathrm{D}}^{\sigma}$.\footnote{It is easy to check that $\{t_{\mathrm{D}}^{\sigma}\,|\,\sigma\in\Lambda\}$ is closed and has a lower bound $t_{\mathrm{cur}}$. Therefore, $\sigma^*$ must exist.} Then
\begin{equation}\label{eqT:prop2}
V(\mathbf{x};t_{\mathrm{cur}},\mathcal{I})=\begin{cases}
C^*(\sigma^*;t_{\mathrm{cur}},\mathcal{I}),&\text{if $\Lambda\neq\varnothing$};\\
+\infty,&\text{otherwise}.
\end{cases}
\end{equation}
Furthermore, if $\Lambda\neq\varnothing$, then $\sigma^*$ has the smallest $t_{0,\mathrm{in}}$ and the largest $v_0(t_{0,\mathrm{in}})$ within all trajectories in $\Sigma_0(\mathbf{x};t_{\mathrm{cur}})$ satisfying \eqref{eqT:safety3}.
\end{proposition}

\begin{IEEEproof}
See Appendix \ref{subsecT:pf_prop2}.
\end{IEEEproof}

\section{Minimax Control Policy}\label{secT:policy}
In this section, we establish the framework of the minimax control policy for the agent $s_0$, present the safety guarantee and provide algorithms for the implementation of the policy.

\subsection{Policy Statement and Safety Guarantee}\label{subsecT:statement}

Recall that decisions are made at times $\{t_k\,|\,k\ge0\}$ where $t_0=0$ and $t_k<t_{k+1}$ for any $k\ge0$. Specifically, we only need to focus on the one-step decision at $t_k$, when the state $\mathbf{x}_0(t_k)=\langle p_0(t_k),v_0(t_k)\rangle$ has been fixed with $p_0(t_k)\le0$, and $s_0$ needs to determine the trajectory during the time interval $[t_k,t_{k+1}]$ based on the situation information $\mathcal{I}_k$.

Firstly, $s_0$ needs to determine whether there exists a trajectory $\sigma\in\Sigma_0(\mathbf{x}_0(t_k);t_k)$ allowing it to safely access the resource during the upcoming interval $[t_k,t_{k+1}]$. In other words, $\sigma$ should simultaneously satisfy
\begin{equation}
(t_{0,\mathrm{in}}^{\sigma},t_{0,\mathrm{out}}^{\sigma})\cap\mathcal{W}(\mathcal{I}_{k}) =\varnothing,\quad
t_{0,\mathrm{in}}^{\sigma}\le t_{k+1}.\label{eqT:cond}
\end{equation}
The next proposition characterizes an equivalent condition for the existence of such a trajectory which is easier to verify.

\begin{proposition}\label{propT:2a}
Let $\Lambda$ be the set of \textsc{Dec}-\textsc{Acc} trajectories on $[t_k,+\infty)$ in $\Sigma_0(\mathbf{x}_0(t_k);t_k)$ satisfying $(t_{0,\mathrm{in}},t_{0,\mathrm{out}})\cap\mathcal{W}(\mathcal{I}_{k}) =\varnothing$. If $\Lambda\neq\varnothing$, let $\sigma^*\triangleq\mathop{\arg\min}_{\sigma\in\Lambda}\;t_{\mathrm{D}}^{\sigma}$. Then a trajectory $\sigma\in\Sigma_0(\mathbf{x}_0(t_k);t_k)$ satisfying \eqref{eqT:cond} exists if and only if
\begin{equation}\label{eqT:prop2a}
\Lambda\neq\varnothing,\quad t_{0,\mathrm{in}}^{\sigma^*}\le t_{k+1}.
\end{equation}
\end{proposition}

\begin{IEEEproof}
If $\sigma^*$ satisfying \eqref{eqT:prop2a} exists, then it can serve as the $\sigma$ we are finding. Conversely, if the trajectory $\sigma$ satisfying \eqref{eqT:cond} exists, then $V(\mathbf{x}_0(t_k);t_k,\mathcal{I}_k)$ is finite. By Proposition \ref{propT:2}, $\Lambda\neq\varnothing$ and we have $t_{0,\mathrm{in}}^{\sigma^*}\le t_{0,\mathrm{in}}^{\sigma}\le t_{k+1}$, which completes the proof.
\end{IEEEproof}

According to Proposition \ref{propT:2a}, if \eqref{eqT:prop2a} holds, then $s_0$ is able to obtain the resource before $t_{k+1}$. In this case, $\sigma^*$ is exactly the trajectory which can result in the smallest manageable cost by Proposition \ref{propT:2}, and thus $s_0$ should follow $\sigma^*$ during $[t_k,t_{k+1}]$. On the contrary, if \eqref{eqT:prop2a} does not hold, we divide the decision of $s_0$ at $t_k$ into two steps. First, a target state $\mathbf{x}_{\mathrm{tar}}=\langle p_{\mathrm{tar}},v_{\mathrm{tar}}\rangle$ is determined which satisfies $p_{\mathrm{tar}}\le0$, and second, a trajectory $\sigma\in\Sigma_0(\mathbf{x}_0(t_k);t_k)$ on the interval $[t_k,t_{k+1}]$ is generated which satisfies
\begin{equation}\label{eqT:step2}
\mathbf{x}_0^{\sigma}(t_{k+1})=\mathbf{x}_{\mathrm{tar}}.
\end{equation}

Let $\mathcal{F}_k$ be the set of target states such that the corresponding trajectory $\sigma$ satisfying \eqref{eqT:step2} exists, i.e.,
\begin{align}
\mathcal{F}_k
&\triangleq\left\{\mathbf{x}_0^{\sigma}(t_{k+1})\,\middle|\,
\begin{aligned}&\sigma\in\Sigma_0(\mathbf{x}_0(t_k);t_k)\\
&p_0^{\sigma}(t_{k+1})\le0\end{aligned}\right\}.\label{eqT:F}
\end{align}
Then we can find that as long as $\mathbf{x}_{\mathrm{tar}}\in\mathcal{F}_k$ is fixed, the choice of the specific trajectory $\sigma$ in the second step does not make any difference for subsequent decisions. Therefore, we focus on the first step of the decision, which aims to determine which target state $\mathbf{x}_{\mathrm{tar}}\in\mathcal{F}_k$ is the most preferred for the time $t_{k+1}$.

Recall that the state value function $V(\mathbf{x}_{\mathrm{tar}};t_{k+1},\mathcal{I}_{k+1})$ proposed in Definition \ref{defT:evalfunc} is a natural criterion to evaluate the preference of different target states. However, note that $s_0$ have no access to $\mathcal{I}_{k+1}$ at the decision time $t_k$. Therefore, a \emph{minimax framework} is adopted for the decision, where we take into account all possible $\mathcal{I}_{k+1}$ and optimize the worst-case performance. Specifically, note that since $\mathcal{I}_{k+1}$ must be consistent with the current information $\mathcal{I}_k$, it must lie in
\begin{equation}\label{eqT:J}
\mathscr{J}_k\triangleq\left\{\mathcal{I}_{k+1}\, \middle|\,\begin{aligned}&\mathcal{I}_{k+1}\subseteq\max\{t_{k+1},\mathcal{I}_{k}\}\\
&\text{satisfying Assumption \ref{asT:2}}\end{aligned}\right\}
\end{equation}
where
\begin{align}
&\max\{t_{k+1},\mathcal{I}_k\}\notag\\
&\quad\triangleq\left\{\begin{aligned}\big\langle&\max\{t_{k+1},\hat{t}_{1,\mathrm{in}}\},\\
&\max\{t_{k+1}, \hat{t}_{1,\mathrm{out}}\}\big\rangle\end{aligned} \,\middle|\,\langle\hat{t}_{1,\mathrm{in}},\hat{t}_{1,\mathrm{out}}\rangle\in\mathcal{I}_k\right\}.
\end{align}
Then the decision is made by
\begin{equation}\label{eqT:opt}
\mathop{\arg\min}_{\mathbf{x}_{\mathrm{tar}}\in\mathcal{F}_k}\; \max_{\mathcal{I}_{k+1}\in\mathscr{J}_k}\; V(\mathbf{x}_{\mathrm{tar}};t_{k+1},\mathcal{I}_{k+1}).
\end{equation}
The full procedure of the proposed control policy at the decision time $t_k$ is concluded in Algorithm \ref{algT:1}.

\begin{algorithm}[t]
  \caption{The decision of $s_0$ at time $t_k$}\label{algT:1}
  \begin{algorithmic}[1]
  \REQUIRE $L_0$, $v_\mathrm{M}$, $a_{0,\mathrm{m}}$, $a_{0,\mathrm{M}}$, $\mathbf{x}_0(t_k)$, $\mathcal{I}_{k}$.
  \ENSURE Trajectory on $[t_k,t_{k+1}]$.
  \STATE Find $\sigma^*$ by Proposition \ref{propT:2}.
  \IF{\eqref{eqT:prop2a} holds}
  \STATE Return $\sigma^*$.
  \ELSE
  \STATE Solve \eqref{eqT:opt} to obtain $\mathbf{x}_{\mathrm{tar}}\in\mathcal{F}_k$.
  \STATE Return an arbitrary trajectory $\sigma\in\Sigma_0(\mathbf{x}_0(t_k);t_k)$ on $[t_k,t_{k+1}]$ satisfying \eqref{eqT:step2}.
  \ENDIF
  \end{algorithmic}
\end{algorithm}

Now we analyze the safety of the policy; in other words, we need to show the output trajectory of the proposed policy satisfies the safety condition \eqref{eqT:safety2}. Recall \eqref{eqT:priorcost} and \eqref{eqT:evalfunc} that for a state $\mathbf{x}$ at time $t_k$, $V(\mathbf{x};t_k,\mathcal{I}_k)<+\infty$ implies that there exists a trajectory in $\Sigma_0(\mathbf{x};t_k)$ satisfying the robust safety condition, which intuitively means that the state $\mathbf{x}$ has the potential of maintaining safety in the future. Therefore, the safety of the policy can be verified by showing that $V(\mathbf{x}_0(t_k);t_k,\mathcal{I}_k)<+\infty$ always holds at each time $t_k$, which can be proved by induction. Specifically, we have the following theorem.

\begin{theorem}\label{thmT:2}
Let $\{t_k\,|\,k\ge0\}$ be a set of decision times of $s_0$ with $t_0=0$ and $t_k<t_{k+1}$ for any $k\ge0$. At each decision time, $s_0$ makes decision according to Algorithm \ref{algT:1}. Then the trajectory of $s_0$ satisfies the safety condition \eqref{eqT:safety2}.
\end{theorem}

\begin{IEEEproof}
See Appendix \ref{subsecT:pf_thm2}.
\end{IEEEproof}

Finally, we provide a remark on the proposed framework.
\begin{remark}
Note that this article focuses on the perspective of one agent instead of the entire system, and agents are not assumed to be cooperative. Therefore, if the two agents both follow the proposed policy, deadlocks may occur since neither of them dare to occupy the resource without knowing how the other agent will react. To eliminate the deadlock, extra mechanisms need to be introduced such as directly specifying the order by a central node when both agents have stopped near the resource.
\end{remark}

The next two subsections provide further characterizations and algorithms on \eqref{eqT:opt}.

\subsection{Range of Feasible Target States}\label{subsecT:feas}
In this subsection, we provide another descriptive characterization of the set $\mathcal{F}_k$ defined in \eqref{eqT:F}, which serves as the objective space of the optimization problem \eqref{eqT:opt}.


\begin{proposition}\label{propT:3}
Fix the state $\mathbf{x}_0(t_k)$. Then
\begin{equation}\label{eqT:tildeF}
\mathcal{F}_k=\left\{\langle p,v\rangle\,\middle|\,\begin{aligned}p&\le0\\
v_0^{\sigma_{\mathrm{D}}}(t_{k+1})\le v&\le v_0^{\sigma_{\mathrm{A}}}(t_{k+1})\\
p_0^{\sigma_{\mathrm{DA}}(v)}(t_{k+1})\le p&\le p_0^{\sigma_{\mathrm{AD}}(v)}(t_{k+1})\end{aligned}\right\}\!\!
\end{equation}
where $\sigma_{\mathrm{A}},\sigma_{\mathrm{D}}\in\Sigma_0(\mathbf{x}_0(t_k);t_k)$ are the \textsc{Acc} trajectory and the \textsc{Dec} trajectory on $[t_k,t_{k+1}]$, respectively; for a given $v$, $\sigma_{\mathrm{DA}}(v),\sigma_{\mathrm{AD}}(v)\in\Sigma_0(\mathbf{x}_0(t_k);t_k)$ are the \textsc{Dec}-\textsc{Acc} trajectory and the \textsc{Acc}-\textsc{Dec} trajectory on $[t_k,t_{k+1}]$ satisfying $v_0^{\sigma_{\mathrm{DA}}(v)}(t_{k+1})=v_0^{\sigma_{\mathrm{AD}}(v)}(t_{k+1})=v$, respectively.\footnote{For $v\in [v_0^{\sigma_{\mathrm{D}}}(t_{k+1}),v_0^{\sigma_{\mathrm{A}}}(t_{k+1})]$, the existence of $\sigma_{\mathrm{DA}}(v)$ (or $\sigma_{\mathrm{AD}}(v)$) comes from the continuity of $v_0^{\sigma}(t_{k+1})$ with respect to $t_{\mathrm{D}}^{\sigma}$ (or $t_{\mathrm{A}}^{\sigma}$), which is discussed in Lemma \ref{lemmaT:2} in Appendix \ref{subsecT:lemma}.}
\end{proposition}

\begin{figure} [!t]
  \centering%
  \includegraphics[width=0.48\textwidth]{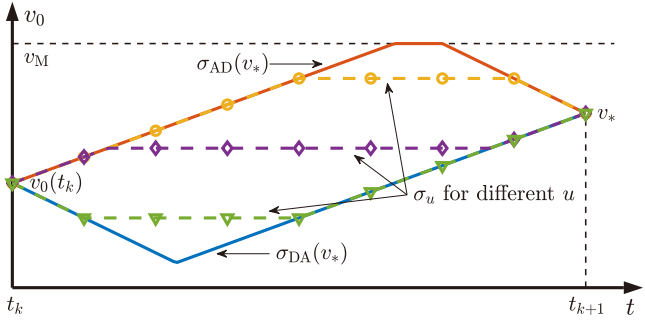}
  \caption{The illustration of the parameterized trajectory family $\{\sigma_u\}$ defined by \eqref{eqT:eg} deforming $\sigma_{\mathrm{DA}}(v_*)$ into $\sigma_{\mathrm{AD}}(v_*)$, where $v_0^{\sigma_u}(t_{k+1})=v_*$ always holds.}
  \label{figT:3}
\end{figure}

\begin{IEEEproof}
For any $\mathbf{x}_0^{\sigma}(t_{k+1})\in\mathcal{F}_k$, we can check that it also belongs to the right-hand side of \eqref{eqT:tildeF}. Conversely, for any $\mathbf{x}_*=\langle p_*,v_*\rangle$ in the right-hand side of \eqref{eqT:tildeF}, we need to find a trajectory $\sigma\in\Sigma_0(\mathbf{x}_0(t_k);t_k)$ with $\mathbf{x}_0^{\sigma}(t_{k+1})=\mathbf{x}_*$. To this end, we construct a one-dimensional parameterized family $\{\sigma_u\}\subseteq \Sigma_0(\mathbf{x}_0(t_k);t_k)$ of trajectories which continuously deforms $\sigma_{\mathrm{DA}}(v_*)$ into $\sigma_{\mathrm{AD}}(v_*)$ with $v_0^{\sigma_u}(t_{k+1})=v_*$ always holds. For example, we can let
\begin{equation}\label{eqT:eg}
v_0^{\sigma_u}(t)=\max\left\{\min\big\{u,v_0^{\sigma_{\mathrm{AD}}(v_*)}(t)\big\},v_0^{\sigma_{\mathrm{DA}}(v_*)}(t)\right\}
\end{equation}
for all $t\in[t_k,t_{k+1}]$, and Fig. \ref{figT:3} provides the illustration of this example on the $(v_0,t)$-plane. Note that by the continuity, the family $\{\sigma_u\}$ must include a trajectory $\sigma_{u_0}$ satisfying $p_0^{\sigma_{u_0}}(t_{k+1})=p_*$. This completes the proof.
\end{IEEEproof}

Note that compared with the original definition \eqref{eqT:F} of $\mathcal{F}_k$, \eqref{eqT:tildeF} is more descriptive since it directly characterizes the inequalities on $p$ and $v$. Furthermore, the proof of Proposition \ref{propT:3} actually provides a method to find a trajectory on $[t_k,t_{k+1}]$ after given the target state $\mathbf{x}_{\mathrm{tar}}\in\mathcal{F}_k$, which can be applied in the Line 6 of Algorithm \ref{algT:1}.

\subsection{Calculating the Maximum}\label{subsecT:cal}
In this subsection, we focus on calculating the objective function of the minimization in \eqref{eqT:opt}, i.e.,
\begin{align}\label{eqT:obj_ori}
&V_{\mathrm{max}}(\mathbf{x}_{\mathrm{tar}};t_{k+1},\mathcal{I}_k)\triangleq\max_{\mathcal{I}_{k+1}\in\mathscr{J}_k}\; V(\mathbf{x}_{\mathrm{tar}};t_{k+1},\mathcal{I}_{k+1})
\end{align}
which itself is represented by another optimization problem. For simplicity of notations, we omit $\mathbf{x}_{\mathrm{tar}}$ and $t_{k+1}$ which are always fixed in this subsection, and write
\begin{equation}\label{eqT:obj}
V_{\mathrm{max}}(\mathcal{I}_k)\triangleq\max_{\mathcal{I}_{k+1}\in\mathscr{J}_k}\; V(\mathcal{I}_{k+1}).
\end{equation}

\begin{figure} [!t]
  \centering%
  \includegraphics[width=0.48\textwidth]{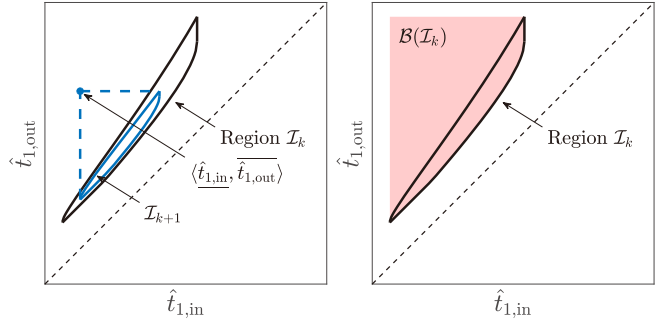}
  \caption{The left subfigure shows the calculation of $\langle\underline{\hat{t}_{1,\mathrm{in}}},\overline{\hat{t}_{1,\mathrm{out}}}\rangle$ for a specific $\mathcal{I}_{k+1}\in\mathscr{J}_k$, and the right subfigure illustrates the set $\mathcal{B}(\mathcal{I}_k)$ defined in Proposition \ref{propT:4} which contains all possible $\langle\underline{\hat{t}_{1,\mathrm{in}}},\overline{\hat{t}_{1,\mathrm{out}}}\rangle$.}
  \label{figT:4}
\end{figure}

According to Definitions \ref{defT:priorcost} and \ref{defT:evalfunc}, the state value function $V(\mathcal{I}_{k+1})$ only depends on $\mathcal{W}(\mathcal{I}_{k+1})$, where the definition of $\mathcal{W}(\cdot)$ is provided in \eqref{eqT:W}. By utilizing Assumption \ref{asT:2}, for any $\mathcal{I}_{k+1}\in\mathscr{J}_k$, we have
\begin{equation}
\mathcal{W}(\mathcal{I}_{k+1})=(\underline{\hat{t}_{1,\mathrm{in}}},\overline{\hat{t}_{1,\mathrm{out}}})\label{eqT:calW}
\end{equation}
where
\begin{align}
\underline{\hat{t}_{1,\mathrm{in}}}&\triangleq\min\,\{\hat{t}_{1,\mathrm{in}}\,|\,\exists\, \hat{t}_{1,\mathrm{out}}, \langle\hat{t}_{1,\mathrm{in}},\hat{t}_{1,\mathrm{out}}\rangle\in\mathcal{I}_{k+1}\}\label{eqT:tm}\\
\overline{\hat{t}_{1,\mathrm{out}}}&\triangleq\max\,\{\hat{t}_{1,\mathrm{out}}\,|\,\exists\, \hat{t}_{1,\mathrm{in}}, \langle\hat{t}_{1,\mathrm{in}},\hat{t}_{1,\mathrm{out}}\rangle\in\mathcal{I}_{k+1}\}.\label{eqT:tM}
\end{align}
The calculation of $\underline{\hat{t}_{1,\mathrm{in}}}$ and $\overline{\hat{t}_{1,\mathrm{out}}}$ is illustrated in the left subfigure in Fig. \ref{figT:4}. Therefore, we can use the alternative notation $V(\underline{\hat{t}_{1,\mathrm{in}}},\overline{\hat{t}_{1,\mathrm{out}}})$ for $V(\mathcal{I}_{k+1})$.

Now we characterize the range of the pair $\langle\underline{\hat{t}_{1,\mathrm{in}}},\overline{\hat{t}_{1,\mathrm{out}}}\rangle$ in the next proposition.

\begin{proposition}\label{propT:4}
For a given situation information $\mathcal{I}_k$, the corresponding range of $\langle\underline{\hat{t}_{1,\mathrm{in}}},\overline{\hat{t}_{1,\mathrm{out}}}\rangle$ is
\begin{align}
\mathcal{B}(\mathcal{I}_k)&\triangleq \Big\{\big\langle\min\{\hat{t}_{1,\mathrm{in}}^{(1)},\hat{t}_{1,\mathrm{in}}^{(2)}\}, \max\{\hat{t}_{1,\mathrm{out}}^{(1)},\hat{t}_{1,\mathrm{out}}^{(2)}\}\big\rangle\,\Big|&\notag\\ &\qquad\qquad\qquad\langle\hat{t}_{1,\mathrm{in}}^{(1)},\hat{t}_{1,\mathrm{out}}^{(1)}\rangle, \langle\hat{t}_{1,\mathrm{in}}^{(2)},\hat{t}_{1,\mathrm{out}}^{(2)}\rangle \in\mathcal{I}_k\Big\}.
\end{align}
\end{proposition}

\begin{IEEEproof}
See Appendix \ref{subsecT:pf_prop4}.
\end{IEEEproof}

The set $\mathcal{B}(\mathcal{I}_k)$ is shown in the right subfigure in Fig. \ref{figT:4}. According to Proposition \ref{propT:4}, our goal of calculating \eqref{eqT:obj} is transformed to
\begin{equation}\label{eqT:obj2}
V_{\mathrm{max}}(\mathcal{I}_k)=
\max_{\langle\underline{t},\overline{t}\rangle\in\mathcal{B}(\mathcal{I}_k)}\; V(\underline{t},\overline{t}).
\end{equation}
Note that here we regard $\underline{t}$ and $\overline{t}$ as two free variables. Before solving \eqref{eqT:obj2}, we first show another property of the optimization range $\mathcal{B}(\mathcal{I}_k)$.

\begin{proposition}\label{propT:4a}
Let $\langle\underline{t}^{(1)},\overline{t}^{(1)}\rangle, \langle\underline{t}^{(2)},\overline{t}^{(2)}\rangle \in\mathcal{B}(\mathcal{I}_k)$. Then the following four pairs are also in $\mathcal{B}(\mathcal{I}_k)$:
\begin{align}
&\big\langle\underline{t}^{(1)}, \max\{\overline{t}^{(1)},\overline{t}^{(2)}\}\big\rangle,\;
&&\big\langle\underline{t}^{(2)}, \max\{\overline{t}^{(1)},\overline{t}^{(2)}\}\big\rangle, \label{eqT:prop4a_1}\\
&\big\langle\min\{\underline{t}^{(1)},\underline{t}^{(2)}\}, \overline{t}^{(1)}\big\rangle,\;
&&\big\langle\min\{\underline{t}^{(1)},\underline{t}^{(2)}\}, \overline{t}^{(2)}\big\rangle.\label{eqT:prop4a_2}
\end{align}
\end{proposition}

\begin{IEEEproof}
See Appendix \ref{subsecT:pf_prop4a}.
\end{IEEEproof}

\begin{figure} [!t]
  \centering%
  \includegraphics[width=0.48\textwidth]{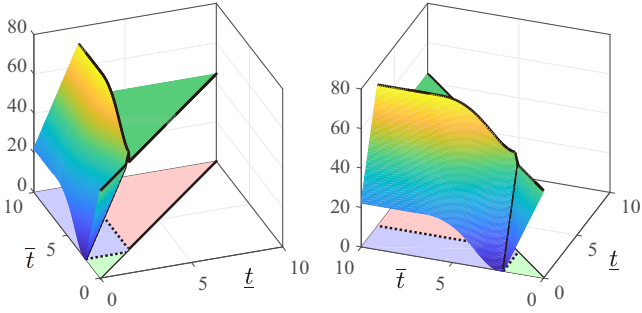}
  \caption{The graph of the function $V(\underline{t},\overline{t})$ from two angles. Parameters $v_{\mathrm{M}}=20$, $a_{0,\mathrm{m}}=4$, $a_{0,\mathrm{M}}=3$, $L_0=5$, $p_{\mathrm{tar}}=-40$, and $v_{\mathrm{tar}}=15$. The domain division is shown on the ground plane that $\mathcal{D}_1$ is colored in red, $\mathcal{D}_2$ in green, and $\mathcal{D}_3$ in blue. The boundaries of the three sub-regions are also displayed on the graph of the function by black curves.}
  \label{figT:5}
\end{figure}

Now we analyze the monotonicity of the function $V(\underline{t},\overline{t})$, which is defined on the domain
\begin{equation}
\mathcal{D}\triangleq\{\langle\underline{t},\overline{t}\rangle\,|\,t_{k+1}\le \underline{t} \le \overline{t} \le B\}.
\end{equation}
To start, we divide this domain into the following three parts\footnote{Although we use the word ``divide'', we set all three sub-regions closed for convenience, and thus their mutual intersections are one-dimensional line segments instead of the empty set.}
\begin{align}
\mathcal{D}_1&\triangleq\big\{\langle\underline{t},\overline{t}\rangle\,\big|\,t_{0,\mathrm{out}}^{\sigma_{\mathrm{A}}}\le \underline{t} \le \overline{t} \le B\big\}\label{eqT:D1}\\
\mathcal{D}_2&\triangleq\big\{\langle\underline{t},\overline{t}\rangle\,\big|\,t_{k+1}\le \underline{t} \le \overline{t} \le t_{0,\mathrm{in}}^{\sigma_{\mathrm{A}}}\big\}\label{eqT:D2}\\
\mathcal{D}_3&\triangleq\left\{\langle\underline{t},\overline{t}\rangle\,\middle|\,\begin{aligned}
t_{k+1}&\le \underline{t} \le t_{0,\mathrm{out}}^{\sigma_{\mathrm{A}}}\\
t_{0,\mathrm{in}}^{\sigma_{\mathrm{A}}}&\le \overline{t} \le B\\
\underline{t} &\le \overline{t}
\end{aligned}\right\}\label{eqT:D3}
\end{align}
in which $\sigma_{\mathrm{A}}\in\Sigma_0(\mathbf{x}_{\mathrm{tar}};t_{k+1})$ is the \textsc{Acc} trajectory on $[t_{k+1},+\infty)$. The next theorem analyzes the monotonicity of $V(\underline{t},\overline{t})$ on the three sub-regions defined above.

\begin{theorem}\label{thmT:1}
For a given state $\mathbf{x}_{\mathrm{tar}}$ of $s_0$ at time $t_{k+1}$, the function $V(\underline{t},\overline{t})$ satisfies the following properties.
\begin{itemize}
\item[(i)] $V$ is constant on $\mathcal{D}_1$.
\item[(ii)] $V$ only depends on $\overline{t}-\underline{t}$ on $\mathcal{D}_2$, and decreases with $\overline{t}-\underline{t}$. Furthermore, the value of $V$ on $\mathcal{D}_2$ is always no larger than its value on $\mathcal{D}_1$.
\item[(iii)] $V$ is non-decreasing with respect to both $\underline{t}$ and $\overline{t}$ on $\mathcal{D}_3$.
\end{itemize}
\end{theorem}

\begin{IEEEproof}
See Appendix \ref{subsecT:pf_thm1}.
\end{IEEEproof}

Fig. \ref{figT:5} provides the graph of the function $V(\underline{t},\overline{t})$ from two angles under certain parameters, which supports our results in Theorem \ref{thmT:1}. Based on the theorem, solving the optimization problem \eqref{eqT:obj2} becomes straightforward, and the solution is presented in the next proposition.

\begin{proposition}\label{corT:4}
Fix the state $\mathbf{x}_{\mathrm{tar}}$ of $s_0$ at time $t_{k+1}$ and the situation information $\mathcal{I}_k$ at time $t_k$.
\begin{itemize}
\item[(i)] If $\mathcal{B}(\mathcal{I}_k)\cap\mathcal{D}_1$ is non-empty, then the maximum value of $V(\underline{t},\overline{t})$ in $\mathcal{B}(\mathcal{I}_k)\cap\mathcal{D}_1$ is obtained at any point in $\mathcal{B}(\mathcal{I}_k)\cap\mathcal{D}_1$.
\item[(ii)] If $\mathcal{B}(\mathcal{I}_k)\cap\mathcal{D}_2$ is non-empty, then the maximum value of $V(\underline{t},\overline{t})$ in $\mathcal{B}(\mathcal{I}_k)\cap\mathcal{D}_2$ is obtained at
    \begin{equation}\label{eqT:cor4_1}
    \bigg\langle\min_{\langle\underline{t},\overline{t}\rangle\in\mathcal{B}(\mathcal{I}_k)\cap\mathcal{D}_2} \underline{t}, \min_{\langle\underline{t},\overline{t}\rangle\in\mathcal{B}(\mathcal{I}_k)\cap\mathcal{D}_2} \overline{t}\bigg\rangle.
    \end{equation}
\item[(iii)] If $\mathcal{B}(\mathcal{I}_k)\cap\mathcal{D}_3$ is non-empty, then the maximum value of $V(\underline{t},\overline{t})$ in $\mathcal{B}(\mathcal{I}_k)\cap\mathcal{D}_3$ is obtained at
    \begin{equation}\label{eqT:cor4_2}
    \bigg\langle\max_{\langle\underline{t},\overline{t}\rangle\in\mathcal{B}(\mathcal{I}_k)\cap\mathcal{D}_3} \underline{t}, \max_{\langle\underline{t},\overline{t}\rangle\in\mathcal{B}(\mathcal{I}_k)\cap\mathcal{D}_3} \overline{t}\bigg\rangle.
    \end{equation}
\end{itemize}
\end{proposition}

\begin{IEEEproof}
See Appendix \ref{subsecT:pf_cor4}.
\end{IEEEproof}

According to Proposition \ref{corT:4}, we only need to calculate \emph{at most three values} of $V(\underline{t},\overline{t})$ to solve \eqref{eqT:obj2}, or equivalently, \eqref{eqT:obj}. This number can be further reduced into two, since if $\mathcal{B}(\mathcal{I}_k)$ has non-empty intersections with all three sub-regions, then $\mathcal{B}(\mathcal{I}_k)\cap\mathcal{D}_2$ can be neglected by Theorem \ref{thmT:1}(ii).

\begin{remark}
If $\underline{t}=t_{k+1}$ for all $\langle \underline{t},\overline{t}\rangle\in\mathcal{B}(\mathcal{I}_k)$, then by Proposition \ref{corT:4}, the maximum value of $V(\underline{t},\overline{t})$ in $\mathcal{B}(\mathcal{I}_k)$ can only be obtained at one of the two endpoints, i.e.,
\begin{equation}
\bigg\langle t_{k+1},\max_{\langle t_{k+1},\overline{t}\rangle\in\mathcal{B}(\mathcal{I}_k)} \overline{t}\bigg\rangle\text{ or }
\bigg\langle t_{k+1},\min_{\langle t_{k+1},\overline{t}\rangle\in\mathcal{B}(\mathcal{I}_k)} \overline{t}\bigg\rangle.
\end{equation}
This special case was presented in our previous work \cite{YanShe:C21}.
\end{remark}

In summary, the algorithm of calculating $V_{\mathrm{max}}(\mathcal{I}_k)$ is concluded in Algorithm \ref{algT:2}.

\begin{algorithm}[t]
  \caption{Solving the maximization in \eqref{eqT:obj}}\label{algT:2}
  \begin{algorithmic}[1]
  \REQUIRE $L_0$, $v_\mathrm{M}$, $a_{0,\mathrm{m}}$, $a_{0,\mathrm{M}}$, $\mathbf{x}_{\mathrm{tar}}$, $\mathcal{I}_{k}$.
  \ENSURE $V_{\mathrm{max}}(\mathcal{I}_k)$.
  \STATE Find $\mathcal{B}(\mathcal{I}_k)$ by Proposition \ref{propT:4}.
  \STATE Separately maximizing $V(\underline{t},\overline{t})$ in the three sub-regions by Proposition \ref{corT:4}, and return the one with the largest value.
  \end{algorithmic}
\end{algorithm}

\section{Discussions}\label{secT:disc}
In this section, we provide intuitions and several generalizations on the proposed minimax scheduling framework.

\subsection{Intuitions}\label{subsecT:intui}
In this subsection, we provide intuitive interpretations on the scheduling problem with kinematic constraints and our minimax solution. Generally speaking, when facing another agent $s_1$ with unknown future trajectory, $s_0$ can choose from two \emph{high-level decisions}:
\begin{itemize}
\item[(i)] trying to occupy the resource before $s_1$, or
\item[(ii)] choosing to start the occupation after $s_1$.
\end{itemize}
Clearly, these two decisions will generate contradictory trajectory planning: under the first decision, $s_0$ tends to maintain the maximum velocity as long as the safety is ensured; under the second decision, $s_0$ should estimate $t_{1,\mathrm{out}}$ in order to start the occupation as early and as fast as possible, and deceleration may be required in early stages.

However, due to the trajectory uncertainty of $s_1$, $s_0$ cannot tell in advance which high-level decision is better. Intuitively, the first choice is clearly more efficient if the trial succeeds; however, if $s_1$ moves too fast, then the failed trial will result in a large ``velocity cost''. In contrast, the second choice can reduce the ``velocity cost'', while it may face a large ``time cost'' if $s_1$ moves too slow. Therefore, no matter which high-level decision is made, it may turn out to be inefficient.

To resolve this difficulty, our framework does not make the high-level decision in advance. Instead, the low-level trajectory planning is directly performed, by which the possibilities of both high-level choices are still preserved. Specifically, in each step, the tradeoff between acceleration and deceleration is obtained by optimizing the potential scheduling cost in the worst case, which includes both the ``time cost'' part and the ``velocity cost'' part. Therefore, it is reasonable for our minimax framework to have a robust scheduling performance.

\subsection{Generalizations}\label{subsecT:gene}
In this subsection, we introduce some generalizations on the applicable scenarios of the proposed scheduling framework.

\subsubsection{Weakly Cooperative Agents}
Several types of weak cooperation between $s_0$ and $s_1$ can be introduced into the proposed framework.\footnote{By ``weak'', we mean that $s_0$ and $s_1$ are still separately controlled.} For example, if $s_1$ makes \emph{promises} to $s_0$ on its velocity (or acceleration) constraint which is stricter than \eqref{eqT:v_cons} (or \eqref{eqT:a_cons}), then $s_0$ can correspondingly change the axis scaling of $s_1$ (or $a_{1,\mathrm{m}}$ and $a_{1,\mathrm{M}}$) in the calculation.

Another commonly-used weak cooperation scheme is determining the \emph{resource occupation order} through communications. If $s_0$ has the priority, then it can ignore $s_1$ and follow the \textsc{Acc} trajectory. Otherwise, if $s_1$ has the priority, then $s_0$ should change the length of $s_1$ to $L_1+\alpha$ and change any position observation $p_1(t_{\mathrm{obs}})$ to $p_1(t_{\mathrm{obs}})+\alpha$ for a large enough $\alpha$ in its calculation.

\subsubsection{Regularity of Information Collection}
In our framework introduced above, we did not assume any regularity on the process of information collection. In other words, at a decision time $t_k$, the uncertainty set for the next decision time $t_{k+1}$ can take any element in $\mathscr{J}_k$ as defined in \eqref{eqT:J}. However, a certain regularity can help us further narrow down the set $\mathscr{J}_k$. For example, if $s_0$ already knows at $t_k$ that it will not obtain any new information at $t_{k+1}$,\footnote{This can occur, for example, if $s_0$ can only make observations at a lower frequency than the decision making, and no observations can be made during $[t_k,t_{k+1}]$.} then the maximum in \eqref{eqT:opt} can be removed by taking $\mathcal{I}_{k+1}=\max\{t_{k+1},\mathcal{I}_k\}$; in contrast, if $s_0$ is confident to make undelayed observations at all decision times,\footnote{This can occur, for example, if the observation frequency of $s_0$ is much higher than than the decision frequency, and the observation delays are much lower than the dynamics of agents.} then $\mathcal{I}_{k+1}$ must lie in a $2$-dimensional subset of $\mathscr{J}_k$ parameterized by the observations. Note that although the framework is always adaptable, the algorithm in Section \ref{subsecT:cal} needs to be re-developed case by case.

\subsubsection{Bounded Control Error}
Assume that there exist bounded control errors for the agent $s_0$. In other words, when $s_0$ tries to control itself to the state $\mathbf{x}_{\mathrm{tar}}$ at time $t_k$, it will actually arrive at the state $\mathbf{x}_{\mathrm{tar}}+\mathbf{n}$ at time $t_{k+1}$, where the unknown error $\mathbf{n}$ is bounded. In this case, we can add a maximization over the control errors and change \eqref{eqT:opt} to
\begin{align}
\min_{\mathbf{x}_{\mathrm{tar}}\in\mathcal{F}_k}\,\max_{\mathbf{n}}\, \max_{\mathcal{I}_{k+1}\in\mathscr{J}_k}V(\mathbf{x}_{\mathrm{tar}}+\mathbf{n};t_{k+1},\mathcal{I}_{k+1}).
\end{align}



\section{Numerical Results}\label{secT:result}
In this section, we provide numerical experiments to show the characteristics of the objective function, verify the performance of the proposed scheduling policy, and evaluate the effects of several parameters.

\subsection{Objective Function $V_{\mathrm{max}}$}\label{subsecT:traj}
\begin{figure} [!t]
  \centering%
  \includegraphics[width=0.48\textwidth]{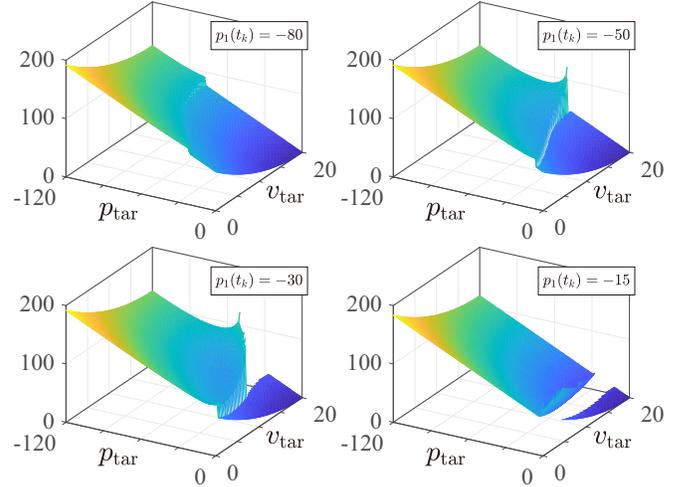}
  \caption{The objective function $V_{\mathrm{max}}(\mathbf{x}_{\mathrm{tar}};t_{k+1},\mathcal{I}_k)$. Parameters $v_{\mathrm{M}}=20$, $a_{0,\mathrm{m}}=a_{1,\mathrm{m}}=4$, $a_{0,\mathrm{M}}=a_{1,\mathrm{M}}=3$, $L_0=L_1=5$, $v_1(t_k)=15$, and $t_{k+1}-t_k=0.01$. Regions with no value represent infinity.}
  \label{figT:simu_1}
\end{figure}

In this subsection, we numerically characterize the objective function $V_{\mathrm{max}}(\mathbf{x}_{\mathrm{tar}};t_{k+1},\mathcal{I}_k)$ of the one-step decision \eqref{eqT:opt} at time $t_k$, which is defined in \eqref{eqT:obj_ori}. We assume that $s_0$ is able to observe $\mathbf{x}_1(t_k)$, based on which the situation information $\mathcal{I}_k$ can be calculated by Proposition \ref{propT:1}. Then the graph of $V_{\mathrm{max}}(\mathbf{x}_{\mathrm{tar}};t_{k+1},\mathcal{I}_k)$ is shown in Fig. \ref{figT:simu_1} with respect to $\mathbf{x}_{\mathrm{tar}}=\langle p_{\mathrm{tar}},v_{\mathrm{tar}}\rangle$ for four different values of $p_1(t_k)$.

According to Fig. \ref{figT:simu_1}, the general trend of $V_{\mathrm{max}}$ decreases with both $p_{\mathrm{tar}}$ and $v_{\mathrm{tar}}$, which is intuitive since a state closer to the resource and with larger velocity is generally more appealing. However, the existence of $s_1$ makes the function discontinuous. Specifically, the right side of the transition has a lower value of $V_{\mathrm{max}}$, since this region corresponds to the case that $s_0$ is always able to occupy the resource \emph{before} $s_1$ by the \textsc{Acc} trajectory. With $p_1(t_k)$ approaching zero, the transition moves right accordingly and the gap becomes larger, since the room for $s_0$ to make adjustment decreases.

\subsection{Performance Comparison}\label{subsecT:perf}
In this subsection, we compare the output trajectory and the scheduling cost of the proposed policy with two other families of policies $\mathrm{Queueing}(d)$ and $\mathrm{Following}(d)$ for $d\ge0$. Specifically,\footnote{Here, $\sigma_{0,\mathrm{A}}$, $\sigma_{0,\mathrm{D}}$ and $\sigma_{1,\mathrm{D}}$ represent the \textsc{Acc} or the \textsc{Dec} trajectory of $s_0$ or $s_1$ on $[t,+\infty)$, respectively.}
\begin{itemize}
\item $\mathrm{Queueing}(d)$ is the policy in which $s_0$ moves with the maximum possible acceleration which can lead to the state $\langle p_0,v_0\rangle=\langle -d,0\rangle$ before it finds an \textsc{Acc} trajectory ensuring robust safety. Mathematically, $a_0(t)=a_{0,\mathrm{A}}(t)$ if $\sigma_{0,\mathrm{A}}$ satisfies \eqref{eqT:safety3} or if $p_0(t)+\frac{v_0^2(t)}{2a_{0,\mathrm{m}}}<-d$ holds; otherwise, $a_0(t)=a_{0,\mathrm{D}}(t)$.
\item $\mathrm{Following}(d)$ is the policy in which $s_0$ moves with the maximum possible acceleration which can ensure $p_0(t)<p_1(t)-L_1-d$ for all $t$ in the future before it finds an \textsc{Acc} trajectory ensuring robust safety. Mathematically, $a_0(t)=a_{0,\mathrm{A}}(t)$ if $\sigma_{0,\mathrm{A}}$ satisfies \eqref{eqT:safety3} or if $p_0^{\sigma_{0,\mathrm{D}}}(t')<p_1^{\sigma_{1,\mathrm{D}}}(t')-L_1-d$ holds for all $t'\ge t$; otherwise, $a_0(t)=a_{0,\mathrm{D}}(t)$.
\end{itemize}
Intuitively, $\mathrm{Queueing}(d)$ and $\mathrm{Following}(d)$ represent typical policies based on the first and the second high-level decision in Section \ref{subsecT:intui}, respectively.\footnote{Precisely speaking, we use the step-wise version of $\mathrm{Queueing}(d)$ and $\mathrm{Following}(d)$ in simulations, while we will not discuss here in detail.}

Now we set the motion of $s_1$ in simulations. Assume that $s_1$ starts from the position $p_1(0)$ and maintains the constant velocity $15$ until it reaches the position $-(v_{\mathrm{F}}^2-15^2)/(2a_{\mathrm{F}})$; then, it changes the acceleration to $a_{\mathrm{F}}$ until it reaches the position $0$ with velocity $v_{\mathrm{F}}$; finally, it follows the constant velocity $v_{\mathrm{F}}$ until it releases the resource. Here, $v_{\mathrm{F}}$ is a free parameter; $a_{\mathrm{F}}\triangleq a_{1,\mathrm{M}}$ if $v_{\mathrm{F}}\ge 15$ and $a_{\mathrm{F}}\triangleq-a_{1,\mathrm{m}}$ if $v_{\mathrm{F}}<15$.
Furthermore, we assume that $s_0$ can always observe the real-time state of $s_1$.

\begin{figure} [!t]
  \centering%
  \includegraphics[width=0.48\textwidth]{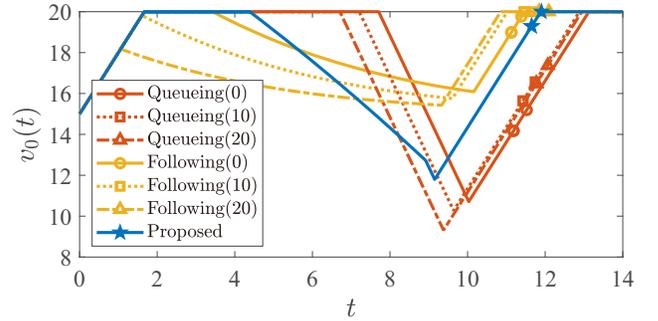}
  \caption{The output trajectories of the proposed policy, $\mathrm{Queueing}(d)$ policies and $\mathrm{Following}(d)$ policies. Parameters $v_{\mathrm{M}}=20$, $a_{0,\mathrm{m}}=a_{1,\mathrm{m}}=4$, $a_{0,\mathrm{M}}=a_{1,\mathrm{M}}=3$, $L_0=L_1=5$, $t_k=0.01k$, $p_0(0)=-200$, $v_0(0)=15$, $p_1(0)=-160$, and $v_{\mathrm{F}}=15$. On each curve, the two markers represent $t=t_{0,\mathrm{in}}$ and $t=t_{0,\mathrm{out}}$.}
  \label{figT:simu_2_traj}
\end{figure}

Fig. \ref{figT:simu_2_traj} provides the output trajectories of different policies for $p_1(0)=-160$ and $v_{\mathrm{F}}=15$. First, we can find that within the $\mathrm{Queueing}(d)$ policies and the $\mathrm{Following}(d)$ policies, the change of $d$ provides different tradeoffs between $t_{0,\mathrm{in}}$ and $v_0(t_{0,\mathrm{in}})$, while the trend of the trajectories in each family is consistent. Second, the output trajectory of the proposed policy intuitively lies \emph{``between''} the $\mathrm{Queueing}(d)$ policies and the $\mathrm{Following}(d)$ policies: the start time of deceleration is later than that of $\mathrm{Following}(d)$, which can preserve the possibility of occupying the resource before $s_1$ when the situation is still unclear; in contrast, it is earlier than that of $\mathrm{Queueing}(d)$, which can avoid a too small occupation velocity. This observation coincides with our intuition in Section \ref{subsecT:intui}. Furthermore, we can also find that the proposed policy turns from deceleration to acceleration \emph{earliest} among all policies.


\begin{figure} [!t]
  \centering%
  \includegraphics[width=0.48\textwidth]{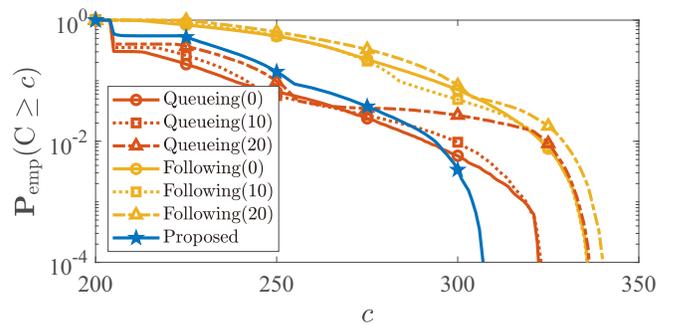}
  \caption{The tail empirical probability $\mathbb{P}_{\mathrm{emp}}(\RS{C}\ge c)$ for the proposed policy, $\mathrm{Queueing}(d)$ policies and $\mathrm{Following}(d)$ policies. Parameters $v_{\mathrm{M}}=20$, $a_{0,\mathrm{m}}=a_{1,\mathrm{m}}=4$, $a_{0,\mathrm{M}}=a_{1,\mathrm{M}}=3$, $L_0=L_1=5$, $t_k=0.01(k-1)$, $p_0(0)=-200$, and $v_0(0)=15$. For each curve, $10^5$ experiments are performed.}
  \label{figT:simu_2_cost}
\end{figure}

Now we perform $10^5$ random experiments by choosing $p_1(0)$ uniformly in $[-200,-100]$ and choosing $v_{\mathrm{F}}$ uniformly in $[5,20]$. The tail empirical probability of the scheduling cost $C$, i.e.,
\begin{equation}
\mathbb{P}_{\mathrm{emp}}(\RS{C}\ge c)\triangleq 10^{-5}\cdot\mathrm{card}\{n\,|\,C^{(n)}\ge c\}
\end{equation}
is shown in Fig. \ref{figT:simu_2_cost}, where $C^{(n)}$ represents the scheduling cost in the $n$-th experiment. As shown in the figure, if we allow an outage probability of $10^{-4}$, the largest cost of the proposed policy is $306.9$, while the largest costs of the other six policies are higher than $322.4$. By discarding the ineliminable constant $204.2$ from both values,\footnote{This ineliminable constant part of the cost results from the initial state $\mathbf{x}_0(0)=\langle-200,15\rangle$ of $s_0$.} this is a reduction by $13.4\%$. Therefore, the simulation result verifies the robustness of the proposed policy against the trajectory uncertainty of $s_1$.
%


\subsection{Effect of Decision and Observation Periods}\label{subsecT:period}
In this subsection, we evaluate the effect of different decision periods and observation periods on the output trajectory. Specifically, the motion of $s_1$ is the same as the last subsection with $p_1(0)=-160$ and $v_{\mathrm{F}}=15$, and we assume that $s_0$ can observe the real-time state of $s_1$ at times $m\delta_{\mathrm{obs}}$ ($m\ge0$). The decision times $t_k$ are set to $k\delta_{\mathrm{dec}}$.

\begin{figure} [!t]
  \centering%
  \includegraphics[width=0.48\textwidth]{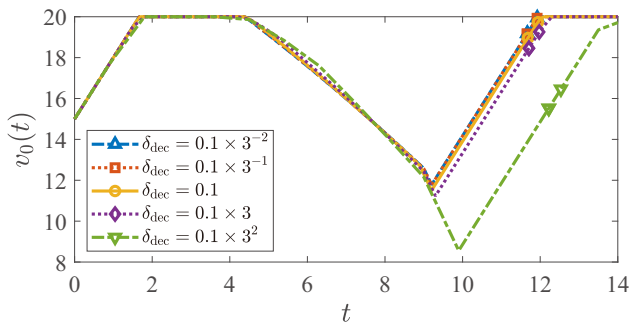}
  \caption{The output trajectories of the proposed policy with different decision periods $\delta_{\mathrm{dec}}$. Parameters $v_{\mathrm{M}}=20$, $a_{0,\mathrm{m}}=a_{1,\mathrm{m}}=4$, $a_{0,\mathrm{M}}=a_{1,\mathrm{M}}=3$, $L_0=L_1=5$, $p_0(0)=-200$, $v_0(0)=15$, $p_1(0)=-160$, $v_{\mathrm{F}}=15$ and $\delta_{\mathrm{obs}}=0.1$. On each curve, the two markers represent $t=t_{0,\mathrm{in}}$ and $t=t_{0,\mathrm{out}}$.}
  \label{figT:simu_3_1}
\end{figure}

\begin{figure} [!t]
  \centering%
  \includegraphics[width=0.48\textwidth]{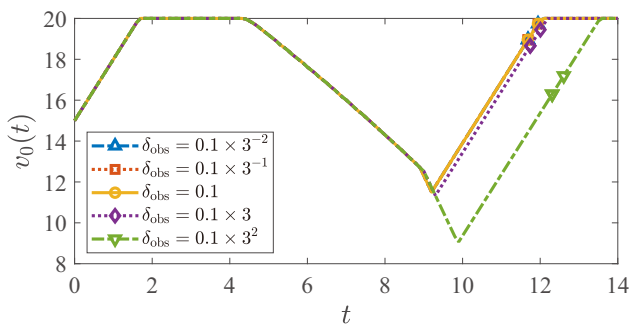}
  \caption{The output trajectories of the proposed policy with different observation periods $\delta_{\mathrm{obs}}$. Parameters $v_{\mathrm{M}}=20$, $a_{0,\mathrm{m}}=a_{1,\mathrm{m}}=4$, $a_{0,\mathrm{M}}=a_{1,\mathrm{M}}=3$, $L_0=L_1=5$, $p_0(0)=-200$, $v_0(0)=15$, $p_1(0)=-160$, $v_{\mathrm{F}}=15$ and $\delta_{\mathrm{dec}}=0.1$. On each curve, the two markers represent $t=t_{0,\mathrm{in}}$ and $t=t_{0,\mathrm{out}}$. Note that the first three curves coincide.}
  \label{figT:simu_3_2}
\end{figure}

Fig. \ref{figT:simu_3_1} and Fig. \ref{figT:simu_3_2} illustrate the output trajectories for different $\delta_{\mathrm{dec}}$ and $\delta_{\mathrm{obs}}$, respectively. According to the two figures, the safety of the system can always be guaranteed, and we can also find that both $\delta_{\mathrm{dec}}$ and $\delta_{\mathrm{obs}}$ have little effect on the performance as long as they are small enough (smaller than $0.3$ under the given parameters). Intuitively, ``small enough'' means that the velocity of $s_1$ should not change much during a period, i.e.,
\begin{equation}
\delta_{\mathrm{dec}},\delta_{\mathrm{obs}}\ll \min\left\{\frac{v_{\mathrm{M}}}{a_{1,\mathrm{m}}},\frac{v_{\mathrm{M}}}{a_{1,\mathrm{M}}}\right\}.
\end{equation}
After that, more frequent decisions or observations will no longer improve the performance.

\subsection{Effect of Different Inertial Constraints}\label{subsecT:iner}
\begin{figure} [!t]
  \centering%
  \includegraphics[width=0.48\textwidth]{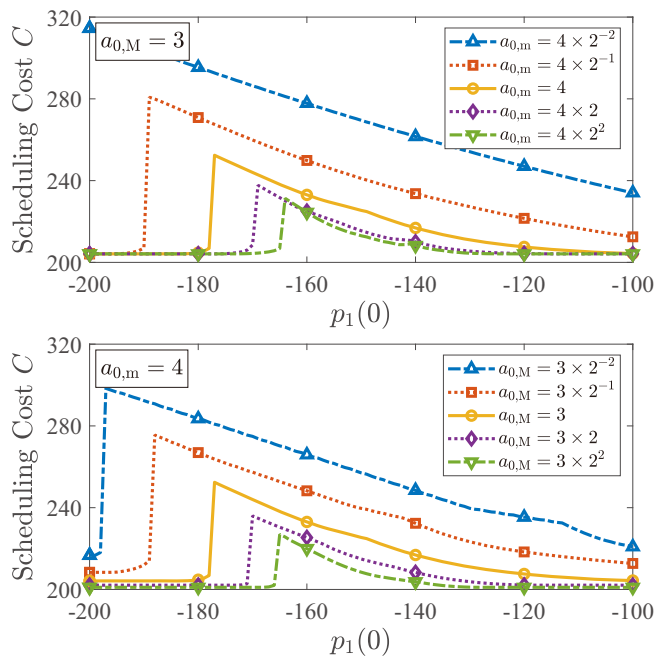}
  \caption{The scheduling cost $C$ of the proposed policy. Parameters $v_{\mathrm{M}}=20$, $a_{1,\mathrm{m}}=4$, $a_{1,\mathrm{M}}=3$, $L_0=L_1=5$, $t_k=0.01(k-1)$, $p_0(0)=-200$, $v_0(0)=15$ and $v_{\mathrm{F}}=15$.}
  \label{figT:simu_4_self}
\end{figure}

\begin{figure} [!t]
  \centering%
  \includegraphics[width=0.48\textwidth]{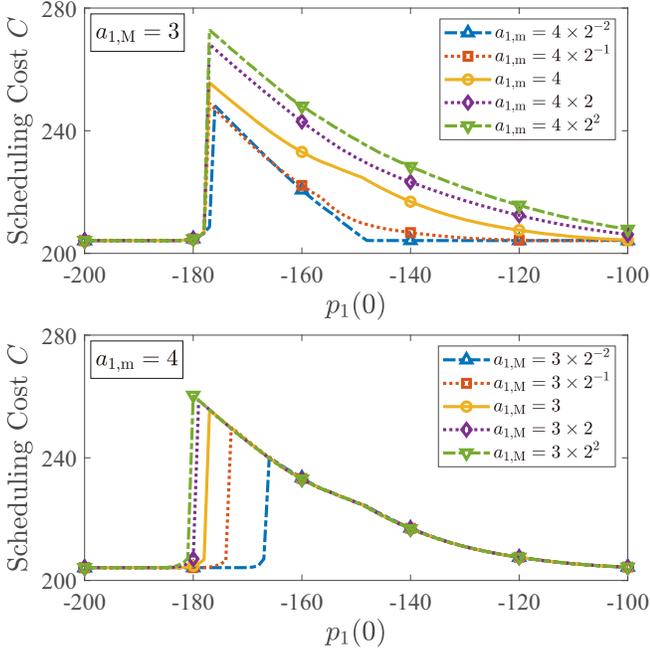}
  \caption{The scheduling cost $C$ of the proposed policy. Parameters $v_{\mathrm{M}}=20$, $a_{0,\mathrm{m}}=4$, $a_{0,\mathrm{M}}=3$, $L_0=L_1=5$, $t_k=0.01(k-1)$, $p_0(0)=-200$, $v_0(0)=15$ and $v_{\mathrm{F}}=15$.}
  \label{figT:simu_4_other}
\end{figure}

In this subsection, we discuss the effect of different inertial constraints of $s_0$ and $s_1$, i.e., the bounds on their accelerations $\langle a_{0,\mathrm{m}},a_{0,\mathrm{M}}\rangle$ and $\langle a_{1,\mathrm{m}},a_{1,\mathrm{M}}\rangle$. For the motion of $s_1$, we still use the one proposed in Section \ref{subsecT:perf} with $v_{\mathrm{F}}=15$. Then the relations between the scheduling cost $C$ and $p_1(0)$ for different $\langle a_{0,\mathrm{m}},a_{0,\mathrm{M}}\rangle$ and $\langle a_{1,\mathrm{m}},a_{1,\mathrm{M}}\rangle$ are shown in Fig. \ref{figT:simu_4_self} and Fig. \ref{figT:simu_4_other}, respectively. Note that all curves are discontinuous: the left (or right) side of the discontinuity means that $s_0$ occupies the resource before (or after) $s_1$.

According to Fig. \ref{figT:simu_4_self}, the changes of $a_{0,\mathrm{m}}$ and $a_{0,\mathrm{M}}$ have similar effects on the scheduling cost, since both of them affect the \emph{control ability} of $s_0$. Clearly, larger $a_{0,\mathrm{m}}$ and $a_{0,\mathrm{M}}$ mean that $s_0$ can adjust the state more free, and thus it can delay the determination on whether to occupy the resource before or after $s_1$ until the situation becomes more clear, which can generally result in a smaller cost.

In contrast, $a_{1,\mathrm{m}}$ and $a_{1,\mathrm{M}}$ is related to the \emph{information collection ability} of $s_0$. As shown in Fig. \ref{figT:simu_4_other}, the effect of $a_{1,\mathrm{m}}$ mainly occurs when $s_0$ occupies the resource after $s_1$. Intuitively, a smaller $a_{1,\mathrm{m}}$ means that $s_0$ can better estimate $t_{1,\mathrm{out}}$ in advance, which earns more time for it to make preparations. On the other hand, the change of $a_{1,\mathrm{M}}$ influences the position of the discontinuity point, since a smaller $a_{1,\mathrm{M}}$ can eliminate possible conflicts caused by the acceleration of $s_1$, which can enable $s_0$ to safely occupy the resource before $s_1$ in some critical cases.

\section{Conclusion}\label{secT:conclu}
This article established a minimax scheduling framework for the system with two inertially constrained agents from the perspective of one agent, where the future trajectory of the other agent is unknown. A control policy was provided where a minimax optimization problem is solved at each decision time, in which the objective function is the state value function for a given situation; the inner maximum is over all possible situations, i.e., the uncertainty sets of the occupation time period of the other agent, for the next decision time; and the outer minimization is over all feasible target states at the next decision time. Calculation methods and the safety guarantee were also presented in the article. Furthermore, numerical simulations were performed comparing the proposed policy with queueing policies and following policies, showing the robustness of our policy against different motion modes of the other agent. In conclusion, the minimax framework provides an effective solution for the scheduling problem of two non-cooperative agents with inertial constraints.

\appendix
In Appendix \ref{subsecT:lemma}, we provide some lemmas to simplify further analyses. Then we present proofs in Appendices \ref{subsecT:pf_prop1}-\ref{subsecT:pf_cor4} for the theoretical results in the article.

\subsection{Lemmas}\label{subsecT:lemma}
The first lemma is a general property in kinematics, which can be directly verified by definition.

\begin{lemma}\label{lemmaT:1}
Let $t_1,t_2\in\mathbb{R}_+$ with $t_1<t_2$. Consider two trajectories $\sigma_1,\sigma_2$ of $s_i$.
\begin{itemize}
\item[(i)] If $p_i^{\sigma_1}(t_1)=p_i^{\sigma_2}(t_1)$, and $v_i^{\sigma_1}(t)\ge v_i^{\sigma_2}(t)$ for all $t\in[t_1,t_2]$, then
    \begin{equation}
    p_i^{\sigma_1}(t_2)\ge p_i^{\sigma_2}(t_2).
    \end{equation}
\item[(ii)] If $p_i^{\sigma_1}(t_1)=p_i^{\sigma_2}(t_1)\le0$, and $v_i^{\sigma_1}(t)\ge v_i^{\sigma_2}(t)$ for all $t\in[t_1,+\infty)$, then
    \begin{equation}
    t_{i,\mathrm{in}}^{\sigma_1}\le t_{i,\mathrm{in}}^{\sigma_2}.
    \end{equation}
\item[(iii)] If $p_i^{\sigma_1}(t_1)=p_i^{\sigma_2}(t_1)\le L_i$, and $v_i^{\sigma_1}(t)\ge v_i^{\sigma_2}(t)$ for all $t\in[t_1,+\infty)$, then
    \begin{equation}
    t_{i,\mathrm{out}}^{\sigma_1}\le t_{i,\mathrm{out}}^{\sigma_2}.
    \end{equation}
\end{itemize}
\end{lemma}

The next lemma characterizes the property of the \textsc{Dec}-\textsc{Acc} trajectories. Similar results for \textsc{Acc}-\textsc{Dec} trajectories also exist by exchanging ``non-decreasing'' and ``non-increasing'', and we leave out detailed expressions.

\begin{lemma}\label{lemmaT:2}
Let $\sigma$ be a \textsc{Dec}-\textsc{Acc} trajectory of $s_i$ on some interval $[t_0,+\infty)$ with a fixed state at $t_0$. Then
\begin{itemize}
\item[(i)] For any fixed $t\ge t_0$, $p_i^{\sigma}(t)$ and $v_i^{\sigma}(t)$ are continuous and non-increasing with respect to $t_{\mathrm{D}}^{\sigma}$;
\item[(ii)] $t_{i,\mathrm{in}}^{\sigma}$ and $t_{i,\mathrm{out}}^{\sigma}$ are continuous and non-decreasing with respect to $t_{\mathrm{D}}^{\sigma}$;
\item[(iii)] $v_i^{\sigma}(t_{i,\mathrm{in}}^{\sigma})$ is non-increasing with respect to $t_{\mathrm{D}}^{\sigma}$.
\end{itemize}
\end{lemma}

\begin{IEEEproof}
(i) Note that the closed-form expression of $p_i^{\sigma}(t)$ and $v_i^{\sigma}(t)$ can be calculated according to Definition \ref{defT:traj}, and the result can be directly checked.

(ii) The monotonicity can be obtained by part (i) and the definitions \eqref{eqT:t_in} and \eqref{eqT:t_out}. The continuity can be verified by separately considering the case $p_i^{\sigma}(t_0)+\frac{v_i^{\sigma}(t_0)^2}{2a_{\mathrm{m}}}>0$ (or $>L$) and the case $p_i^{\sigma}(t_0)+\frac{v_i^{\sigma}(t_0)^2}{2a_{\mathrm{m}}}\le0$ (or $\le L$), which is also straightforward.

(iii) Let $\sigma_1$ and $\sigma_2$ be two \textsc{Dec}-\textsc{Acc} trajectories with $t_{\mathrm{D}}^{\sigma_1}<t_{\mathrm{D}}^{\sigma_2}$, and thus we have
\begin{equation}
v_i^{\sigma_1}(t_{\mathrm{D}}^{\sigma_1})\ge v_i^{\sigma_2}(t_{\mathrm{D}}^{\sigma_2}).
\end{equation}
Now we prove by contradiction. Assume that
\begin{equation}
v_i^{\sigma_1}(t_{i,\mathrm{in}}^{\sigma_1})<v_i^{\sigma_2}(t_{i,\mathrm{in}}^{\sigma_2})
\end{equation}
then we must have
\begin{equation}
t_{i,\mathrm{in}}^{\sigma_2}-t_{\mathrm{D}}^{\sigma_2}>t_{i,\mathrm{in}}^{\sigma_1}-t_{\mathrm{D}}^{\sigma_1}\triangleq\Delta
\end{equation}
which yields
\begin{equation}
v_i^{\sigma_1}(t_{i,\mathrm{in}}^{\sigma_1}-\delta)<v_i^{\sigma_2}(t_{i,\mathrm{in}}^{\sigma_2}-\delta),\;\forall\delta\in[0,\Delta].
\end{equation}
Therefore,
\begin{align}
0&=p_i^{\sigma_1}(t_{i,\mathrm{in}}^{\sigma_1})\notag\\
&=\big(p_i^{\sigma_1}(t_{i,\mathrm{in}}^{\sigma_1})-p_i^{\sigma_1}(t_{\mathrm{D}}^{\sigma_1})\big) +p_i^{\sigma_1}(t_{\mathrm{D}}^{\sigma_1})\notag\\
&<\big(p_i^{\sigma_2}(t_{i,\mathrm{in}}^{\sigma_2})-p_i^{\sigma_2}(t_{i,\mathrm{in}}^{\sigma_2}-\Delta)\big) +p_i^{\sigma_2}(t_{\mathrm{D}}^{\sigma_1})\notag\\
&\le p_i^{\sigma_2}(t_{i,\mathrm{in}}^{\sigma_2})=0
\end{align}
which is a contradiction.
\end{IEEEproof}

The next lemma claims that among all trajectories with the same $t_{i,\mathrm{in}}$, the \textsc{Dec}-\textsc{Acc} trajectory serves as the upper bound for $v_i(t_{i,\mathrm{in}})$. Similarly, the \textsc{Acc}-\textsc{Dec} trajectory serves as the lower bound, and we leave out detailed expressions.

\begin{lemma}\label{lemmaT:3}
Let $\sigma$ be a \textsc{Dec}-\textsc{Acc} trajectory of $s_i$ on some interval $[t_0,+\infty)$. Then for any trajectory $\sigma'$ with the same initial state at $t_0$ and satisfying $t_{i,\mathrm{in}}^{\sigma'}=t_{i,\mathrm{in}}^{\sigma}\triangleq t_{\mathrm{in}}$, we have
\begin{equation}
v_i^{\sigma'}(t_{\mathrm{in}})\le v_i^{\sigma}(t_{\mathrm{in}}).
\end{equation}
\end{lemma}

\begin{IEEEproof}
By the continuity in Lemma \ref{lemmaT:2}(i), there exists another \textsc{Dec}-\textsc{Acc} trajectory $\sigma_1$ such that
\begin{equation}\label{eqT:lemma3_1}
v_i^{\sigma_1}(t_{\mathrm{in}})=v_i^{\sigma'}(t_{\mathrm{in}}).
\end{equation}
Note that by the definition of the \textsc{Dec}-\textsc{Acc} trajectory, for any $t\in[t_{\mathrm{cur}},t_{\mathrm{in}}]$, we have
\begin{align}
v_i^{\sigma'}(t)&\ge\max\big\{v_i^{\sigma'}(t_{\mathrm{cur}})-a_{\mathrm{m}}(t-t_{\mathrm{cur}}),\notag\\ &\phantom{\ge\max\big\{\;}v_i^{\sigma'}(t_{\mathrm{in}})-a_{\mathrm{M}}(t_{\mathrm{in}}-t),0\big\}\notag\\
&=v_i^{\sigma_1}(t).
\end{align}
According to Lemma \ref{lemmaT:1}, we have
\begin{equation}
p_i^{\sigma_1}(t_{\mathrm{in}})\le p_i^{\sigma'}(t_{\mathrm{in}})=0=p_i^{\sigma}(t_{\mathrm{in}})
\end{equation}
and then by Lemma \ref{lemmaT:2}(i), we have $t_{\mathrm{D}}^{\sigma_1}\ge t_{\mathrm{D}}^{\sigma}$ and
\begin{equation}\label{eqT:lemma3_2}
v_i^{\sigma_1}(t_{\mathrm{in}})\le v_i^{\sigma}(t_{\mathrm{in}}).
\end{equation}
By \eqref{eqT:lemma3_1} and \eqref{eqT:lemma3_2}, the proof is completed.
\end{IEEEproof}

\subsection{Proof of Proposition \ref{propT:1} in Section \ref{subsecT:extrac}}\label{subsecT:pf_prop1}
\begin{IEEEproof}
(i) For any trajectory $\sigma$ of $s_1$, $t_{1,\mathrm{in}}^{\sigma_{\mathrm{A}}}\le t_{1,\mathrm{in}}^{\sigma}\le t_{1,\mathrm{in}}^{\sigma_{\mathrm{D}}}$ holds by Lemma \ref{lemmaT:1}(ii). According to Lemma \ref{lemmaT:3}, we have
\begin{equation}
v_1^{\sigma_{\mathrm{AD}}(t_{1,\mathrm{in}}^{\sigma})}(t_{1,\mathrm{in}}^{\sigma})\le v_1^{\sigma}(t_{1,\mathrm{in}}^{\sigma})\le v_1^{\sigma_{\mathrm{DA}}(t_{1,\mathrm{in}}^{\sigma})}(t_{1,\mathrm{in}}^{\sigma}).
\end{equation}
Then by Lemma \ref{lemmaT:1}(iii),
\begin{equation}
t_{1,\mathrm{out}}^{\sigma_{\mathrm{DA}}(t_{1,\mathrm{in}}^{\sigma})}\le t_{1,\mathrm{out}}^{\sigma}\le t_{1,\mathrm{out}}^{\sigma_{\mathrm{AD}}(t_{1,\mathrm{in}}^{\sigma})}.
\end{equation}
Therefore, $\langle\hat{t}_{1,\mathrm{in}}^{\sigma},\hat{t}_{1,\mathrm{out}}^{\sigma}\rangle\in\mathcal{I}$.

(ii) For any trajectory $\sigma$ of $s_1$, $t_{1,\mathrm{out}}^{\sigma_{\mathrm{A}}}\le t_{1,\mathrm{out}}^{\sigma}\le t_{1,\mathrm{out}}^{\sigma_{\mathrm{D}}}$ holds by Lemma \ref{lemmaT:1}(iii). Therefore, $\langle\hat{t}_{1,\mathrm{in}}^{\sigma},\hat{t}_{1,\mathrm{out}}^{\sigma}\rangle\in\mathcal{I}$.

(iii) The result is obvious.
\end{IEEEproof}

\subsection{Proof of Proposition \ref{propT:1a} in Section \ref{subsecT:extrac}}\label{subsecT:pf_prop1a}
\begin{IEEEproof}
The claims in the first sentence are clear. For the monotonicity, we only need to consider the case (i) in Proposition \ref{propT:1}. A further observation is that we only need to show that $t_{1,\mathrm{out}}^{\sigma_{\mathrm{DA}}(t_{1,\mathrm{in}})}-t_{1,\mathrm{in}}$ and $t_{1,\mathrm{out}}^{\sigma_{\mathrm{AD}}(t_{1,\mathrm{in}})}-t_{1,\mathrm{in}}$ are non-decreasing with respect to $t_{1,\mathrm{in}}$.

Let $t_{1,\mathrm{in}}^{(1)}\le t_{1,\mathrm{in}}^{(2)}$. By Lemma \ref{lemmaT:2}(ii), we have
\begin{equation}
t_{\mathrm{D}}^{\sigma_{\mathrm{DA}}(t_{1,\mathrm{in}}^{(1)})}\le t_{\mathrm{D}}^{\sigma_{\mathrm{DA}}(t_{1,\mathrm{in}}^{(2)})}
\end{equation}
and by Lemma \ref{lemmaT:2}(iii),
\begin{equation}
v_1^{\sigma_{\mathrm{DA}}(t_{1,\mathrm{in}}^{(1)})}(t_{1,\mathrm{in}}^{(1)})\ge v_1^{\sigma_{\mathrm{DA}}(t_{1,\mathrm{in}}^{(2)})}(t_{1,\mathrm{in}}^{(2)})
\end{equation}
Then according to Lemma \ref{lemmaT:1}(iii), we have
\begin{equation}
t_{1,\mathrm{out}}^{\sigma_{\mathrm{DA}}(t_{1,\mathrm{in}}^{(1)})}-t_{1,\mathrm{in}}^{(1)}\le t_{1,\mathrm{out}}^{\sigma_{\mathrm{DA}}(t_{1,\mathrm{in}}^{(2)})}-t_{1,\mathrm{in}}^{(2)}.
\end{equation}
Similarly, the monotonicity of $t_{1,\mathrm{out}}^{\sigma_{\mathrm{AD}}(t_{1,\mathrm{in}})}-t_{1,\mathrm{in}}$ can also be obtained.
\end{IEEEproof}

\subsection{Proof of Proposition \ref{propT:2} in Section \ref{subsecT:obj}}\label{subsecT:pf_prop2}
\begin{IEEEproof}
The proof of \eqref{eqT:prop2} is divided into two steps. First, we show that for any trajectory of $s_0$ in $\Sigma_0(\mathbf{x};t_{\mathrm{cur}})$ with finite manageable cost, we can find a \textsc{Dec}-\textsc{Acc} trajectory with an equal or smaller manageable cost. Then, we show that within the \textsc{Dec}-\textsc{Acc} trajectories, the trajectory $\sigma^*$ has the smallest manageable cost.

For the first step, let $\sigma_0\in\Sigma_0(\mathbf{x};t_{\mathrm{cur}})$ be a general trajectory with finite $C^*(\sigma_0;\mathcal{I})$. Then by the continuity in Lemma \ref{lemmaT:2}(ii), there exists a \textsc{Dec}-\textsc{Acc} trajectory $\sigma$ such that $t_{0,\mathrm{in}}^{\sigma}=t_{0,\mathrm{in}}^{\sigma_0}$. Then by Lemma \ref{lemmaT:3}, we have
\begin{equation}
v_0^{\sigma}(t_{0,\mathrm{in}}^{\sigma})\ge v_0^{\sigma_0}(t_{0,\mathrm{in}}^{\sigma_0}).
\end{equation}
Therefore, $C^*(\sigma;\mathcal{I})\le C^*(\sigma_0;\mathcal{I})$, and $\sigma$ is the trajectory we want for the first step.

The second step is the direct corollary of Lemma \ref{lemmaT:2}(ii) and (iii) that for a \textsc{Dec}-\textsc{Acc} trajectory $\sigma$, $t_{0,\mathrm{in}}^{\sigma}$ is non-decreasing with $t_{\mathrm{D}}^{\sigma}$ and $v_0^{\sigma}(t_{0,\mathrm{in}}^{\sigma})$ is non-increasing with $t_{\mathrm{D}}^{\sigma}$. Thus \eqref{eqT:prop2} is proved.

Note that the last result in Proposition \ref{propT:2} is also included in the discussions above. Therefore, the proof is completed.
\end{IEEEproof}

\subsection{Proof of Theorem \ref{thmT:2} in Section \ref{subsecT:statement}}\label{subsecT:pf_thm2}
\begin{IEEEproof}
We denote the final trajectory of $s_0$ by $\sigma_*$, and assume that $t_K<t_{0,\mathrm{in}}^{\sigma_*}\le t_{K+1}$. We also assume that the set of target states at $t_k$ is $\mathcal{F}_{k,*}$, the information obtained at $t_k$ is $\mathcal{I}_{k,*}$, and the range of $\mathcal{I}_{k+1}$ based on $\mathcal{I}_{k,*}$ is $\mathscr{J}_{k,*}$. For $0\le k\le K-1$, we define the following statement:
\begin{align}
\mathscr{P}_k: &\,\exists\, \mathbf{x}_{\mathrm{tar}}\in\mathcal{F}_{k,*},\,
V_{\mathrm{max}}(\mathbf{x}_{\mathrm{tar}};t_{k+1},\mathcal{I}_{k,*})<+\infty.
\end{align}
by the definitions \eqref{eqT:obj_ori} and \eqref{eqT:evalfunc}, $\mathscr{P}_k$ is equivalent to
\begin{align}
\mathscr{P}_k': &\,\exists\,\mathbf{x}_{\mathrm{tar}}\in\mathcal{F}_{k,*},\,
\forall\,\mathcal{I}_{k+1}\in\mathscr{J}_{k,*},\notag\\
&\,\exists\,\sigma\in\Sigma_0(\mathbf{x}_{\mathrm{tar}};t_{k+1}),\,
C^*(\sigma;t_{k+1},\mathcal{I}_{k+1})<+\infty.\!\!
\end{align}
By exchanging the order, the next statement is a sufficient condition for $\mathscr{P}_k'$:
\begin{align}
\mathscr{Q}_k: &\,\exists\,\mathbf{x}_{\mathrm{tar}}\in\mathcal{F}_{k,*},\,
\exists\,\sigma\in\Sigma_0(\mathbf{x}_{\mathrm{tar}};t_{k+1}),\notag\\
&\,\forall\,\mathcal{I}_{k+1}\in\mathscr{J}_{k,*},\, C^*(\sigma;t_{k+1},\mathcal{I}_{k+1})<+\infty.
\end{align}
Then according to \eqref{eqT:priorcost} and \eqref{eqT:evalfunc}, we can successively check that $\mathscr{Q}_k$ is equivalent to the three statements below:
\begin{align}
\mathscr{Q}_k': &\,\exists\,\mathbf{x}_{\mathrm{tar}}\in\mathcal{F}_{k,*},\,
\exists\,\sigma\in\Sigma_0(\mathbf{x}_{\mathrm{tar}};t_{k+1}),\notag\\
&\qquad C^*\big(\sigma;t_{k+1},\max\{t_{k+1},\mathcal{I}_{k,*}\}\big)<+\infty.\\
\mathscr{Q}_k'': &\,\exists\,\sigma\in\Sigma_0(\mathbf{x}_0^{\sigma_*}(t_k);t_k),\,C^*(\sigma;t_k,\mathcal{I}_{k,*})<+\infty.\\
\mathscr{Q}_k''': &\,V(\mathbf{x}_0^{\sigma_*}(t_k);t_k,\mathcal{I}_{k,*})<+\infty.
\end{align}
The definition of $\mathscr{Q}_k''$ and $\mathscr{Q}_k'''$ can also be extended to $k=K$.

Note that $\mathscr{Q}_0''$ is true by \eqref{eqT:initial}, and we have the logic chain
\begin{equation}
\mathscr{Q}_k''\Rightarrow\mathscr{P}_k\Rightarrow\mathscr{Q}_{k+1}'''\Leftrightarrow\mathscr{Q}_{k+1}''
\end{equation}
for $k\le K-1$, where the second inference comes from the choice of $\mathbf{x}_{\mathrm{tar}}$ in Algorithm \ref{algT:1}. Then by induction, $\mathscr{Q}_K''$ is true.
Furthermore, note that when $\mathscr{Q}_K''$ is true, Algorithm \ref{algT:1} can always return a trajectory satisfying the robust safety condition at time $t_K$, which is stronger than the safety condition \eqref{eqT:safety2}. Thus we complete the proof.
\end{IEEEproof}

\subsection{Proof of Proposition \ref{propT:4} in Section \ref{subsecT:cal}}\label{subsecT:pf_prop4}
\begin{IEEEproof}
First, for a given $\mathcal{I}_{k+1}\subseteq\mathcal{I}_k$, there exist two points $\langle\underline{\hat{t}_{1,\mathrm{in}}},t_*\rangle$ and $\langle t_{**},\overline{\hat{t}_{1,\mathrm{out}}}\rangle$ in $\mathcal{I}_{k+1}$ by \eqref{eqT:tm} and \eqref{eqT:tM}. In addition, by the minimality of $\underline{\hat{t}_{1,\mathrm{in}}}$ and the maximality of $\overline{\hat{t}_{1,\mathrm{out}}}$, we have
\begin{equation}
\underline{\hat{t}_{1,\mathrm{in}}}\le t_{**},\quad \overline{\hat{t}_{1,\mathrm{out}}}\ge t_*.
\end{equation}
Therefore, $\langle\underline{\hat{t}_{1,\mathrm{in}}},\overline{\hat{t}_{1,\mathrm{out}}}\rangle\in \mathcal{B}(\mathcal{I}_k)$.

\begin{figure} [!t]
  \centering%
  \includegraphics[width=0.24\textwidth]{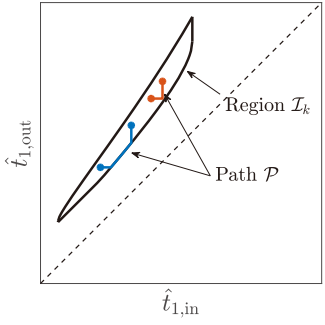}
  \caption{Illustration of the constructions of the path $\mathcal{P}$ in the proof of Proposition \ref{propT:4} for two pairs of points, which correspond to \eqref{eqT:P1} (blue) and \eqref{eqT:P2} (red), respectively.}
  \label{figT:apx}
\end{figure}

Conversely, let $\langle\hat{t}_{1,\mathrm{in}}^{(1)},\hat{t}_{1,\mathrm{out}}^{(1)}\rangle, \langle\hat{t}_{1,\mathrm{in}}^{(2)},\hat{t}_{1,\mathrm{out}}^{(2)}\rangle \in\mathcal{I}_k$, and we show that there exists $\mathcal{I}_{k+1}\subseteq\mathcal{I}_k$ satisfying Assumption \ref{asT:2} such that
\begin{equation}\label{eqT:prop4_1}
\langle\underline{\hat{t}_{1,\mathrm{in}}},\overline{\hat{t}_{1,\mathrm{out}}}\rangle
=\big\langle\min\{\hat{t}_{1,\mathrm{in}}^{(1)},\hat{t}_{1,\mathrm{in}}^{(2)}\},
\max\{\hat{t}_{1,\mathrm{out}}^{(1)},\hat{t}_{1,\mathrm{out}}^{(2)}\}\big\rangle.\!
\end{equation}
Now we separately consider the following two cases.
\begin{itemize}
\item If the right-hand side of \eqref{eqT:prop4_1} equals either $\langle\hat{t}_{1,\mathrm{in}}^{(1)},\hat{t}_{1,\mathrm{out}}^{(1)}\rangle$ or $\langle\hat{t}_{1,\mathrm{in}}^{(2)},\hat{t}_{1,\mathrm{out}}^{(2)}\rangle$, then $\mathcal{I}_{k+1}$ can be chosen as the corresponding single-point set, which clearly satisfies Assumption \ref{asT:2}.
\item Otherwise, we assume that the right-hand side of \eqref{eqT:prop4_1} is equal to $\langle\hat{t}_{1,\mathrm{in}}^{(1)},\hat{t}_{1,\mathrm{out}}^{(2)}\rangle$ without loss of generality. Consider a path $\mathcal{P}$ connecting $\langle\hat{t}_{\mathrm{in}}^{(1)},\hat{t}_{\mathrm{out}}^{(1)}\rangle$ and $\langle\hat{t}_{1,\mathrm{in}}^{(2)},\hat{t}_{1,\mathrm{out}}^{(2)}\rangle$, which is defined as follows: if $\overline{\hat{t}_{1,\mathrm{in}}}(\hat{t}_{1,\mathrm{out}}^{(1)})\le\hat{t}_{1,\mathrm{in}}^{(2)}$, then
    \begin{align}
    \mathcal{P}&\triangleq\left\{\langle t_*,\hat{t}_{1,\mathrm{out}}^{(1)}\rangle\,\middle|\,\hat{t}_{1,\mathrm{in}}^{(1)}\le t_* \le \overline{\hat{t}_{1,\mathrm{in}}}(\hat{t}_{1,\mathrm{out}}^{(1)})\right\}\notag\\
    &\quad\cup\left\{\big\langle t_*,\underline{\hat{t}_{1,\mathrm{out}}}(t_*)\big\rangle\,\middle|\, \overline{\hat{t}_{1,\mathrm{in}}}(\hat{t}_{1,\mathrm{out}}^{(1)})\le t_* \le \hat{t}_{1,\mathrm{in}}^{(2)}\right\}\notag\\
    &\quad\cup\left\{\langle\hat{t}_{1,\mathrm{in}}^{(2)},t_*\rangle\,\middle|\, \underline{\hat{t}_{1,\mathrm{out}}}(\hat{t}_{1,\mathrm{in}}^{(2)})\le t_* \le \hat{t}_{1,\mathrm{out}}^{(2)}\right\}\label{eqT:P1}
    \end{align}
    and otherwise
    \begin{align}
    \mathcal{P}&\triangleq\left\{\langle t_*,\hat{t}_{1,\mathrm{out}}^{(1)}\rangle\,\middle|\,\hat{t}_{1,\mathrm{in}}^{(1)}\le t_* \le \hat{t}_{1,\mathrm{in}}^{(2)}\right\}\notag\\
    &\quad\cup\left\{\langle\hat{t}_{1,\mathrm{in}}^{(2)},t_*\rangle\,\middle|\, \hat{t}_{1,\mathrm{out}}^{(1)}\le t_* \le \hat{t}_{1,\mathrm{out}}^{(2)}\right\}\label{eqT:P2}
    \end{align}
    where
    \begin{align}
    \overline{\hat{t}_{1,\mathrm{in}}}(\cdot)&\triangleq \max\left\{t_*\,\middle|\,\langle t_*,\cdot\rangle\in\mathcal{I}_k\right\}\\
    \underline{\hat{t}_{1,\mathrm{out}}}(\cdot)&\triangleq \min\left\{t_*\,\middle|\,\langle\cdot,t_*\rangle\in\mathcal{I}_k\right\}.
    \end{align}
    The construction of the path $\mathcal{P}$ is shown in Fig. \ref{figT:apx}. By using Assumption \ref{asT:2} on $\mathcal{I}_k$, we can show that $\mathcal{P}$ also satisfies Assumption \ref{asT:2}. Therefore, by choosing $\mathcal{I}_{k+1}=\mathcal{P}$, \eqref{eqT:prop4_1} holds.
\end{itemize}
Thus we complete the proof.
\end{IEEEproof}

\subsection{Proof of Proposition \ref{propT:4a} in Section \ref{subsecT:cal}}\label{subsecT:pf_prop4a}
\begin{IEEEproof}
For \eqref{eqT:prop4a_1}, assume $\overline{t}^{(1)}\ge \overline{t}^{(2)}$ without loss of generality. Then we only need to show $\langle\underline{t}^{(2)},\overline{t}^{(1)}\rangle\in\mathcal{B}(\mathcal{I}_k)$.

Since $\langle\underline{t}^{(1)},\overline{t}^{(1)}\rangle, \langle\underline{t}^{(2)},\overline{t}^{(2)}\rangle\in\mathcal{B}(\mathcal{I}_k)$, there must exist $t_*\ge\underline{t}^{(1)}$ and $t_{**}\le\overline{t}^{(2)}$ such that $\langle t_*,\overline{t}^{(1)}\rangle, \langle\underline{t}^{(2)},t_{**}\rangle \in\mathcal{I}_k$. Clearly, we have
\begin{equation}
t_{**}\le \overline{t}^{(2)} \le \overline{t}^{(1)}.
\end{equation}

Now we choose
\begin{equation}
t_*=\max\big\{t\,\big|\,\langle t,\overline{t}^{(1)}\rangle\in\mathcal{I}_k\big\}\label{eqT:prop4a_3}
\end{equation}
and we aim to show $t_*\ge \underline{t}^{(2)}$ by contradiction. First, let $\underline{\hat{t}_{1,\mathrm{out}}}(\cdot)$ and $\overline{\hat{t}_{1,\mathrm{out}}}(\cdot)$ be defined by
\begin{equation}
\big[\underline{\hat{t}_{1,\mathrm{out}}}(t),\overline{\hat{t}_{1,\mathrm{out}}}(t)\big] \triangleq\mathcal{I}_k\cap\big\{\hat{t}_{1,\mathrm{in}}=t\big\}
\end{equation}
if it is non-empty. By $\langle t_*,\overline{t}^{(1)}\rangle, \langle\underline{t}^{(2)},t_{**}\rangle \in\mathcal{I}_k$, both $\underline{\hat{t}_{1,\mathrm{out}}}(\cdot)$ and $\overline{\hat{t}_{1,\mathrm{out}}}(\cdot)$ are well-defined at $t_*$ and $\underline{t}^{(2)}$. Suppose $t_*<\underline{t}^{(2)}$, and then by Assumption \ref{asT:2}, we have
\begin{equation}
\overline{t}^{(1)}\le \overline{\hat{t}_{1,\mathrm{out}}}(t_*)\le \overline{\hat{t}_{1,\mathrm{out}}}(\underline{t}^{(2)}).
\end{equation}
Since we also have $\underline{\hat{t}_{1,\mathrm{out}}}(\underline{t}^{(2)})\le \overline{t}^{(2)}\le \overline{t}^{(1)}$, it can be concluded that $(\underline{t}^{(2)},\overline{t}^{(1)})\in\mathcal{I}_k$, which contradicts with \eqref{eqT:prop4a_3}. Hence, $t_*\ge \underline{t}^{(2)}$.

Finally, by $\langle t_*,\overline{t}^{(1)}\rangle,\langle\underline{t}^{(2)},t_{**}\rangle\in\mathcal{I}_k$, $t_{**}\le\overline{t}^{(1)}$ and $t_*\ge \underline{t}^{(2)}$, we have $\langle\underline{t}^{(2)},\overline{t}^{(1)}\rangle\in\mathcal{B}(\mathcal{I}_k)$ according to Proposition \ref{propT:4}.

Another result \eqref{eqT:prop4a_2} can be similarly proved.
\end{IEEEproof}

\subsection{Proof of Theorem \ref{thmT:1} in Section \ref{subsecT:cal}}\label{subsecT:pf_thm1}
\begin{IEEEproof}
(i) On $\mathcal{D}_1$, we have $t_{0,\mathrm{out}}^{\sigma_{\mathrm{A}}}\le \underline{t}$. Recall that the interval $(\underline{t},\overline{t})$ corresponds to $\mathcal{W}(\mathcal{I}_{k+1})$ according to \eqref{eqT:calW}. Therefore, $\sigma_{\mathrm{A}}$ satisfies the robust safety condition $(t_{0,\mathrm{in}}^{\sigma_{\mathrm{A}}},t_{0,\mathrm{out}}^{\sigma_{\mathrm{A}}}) \cap\mathcal{W}(\mathcal{I}_{k+1})=\varnothing$.
Since $\sigma_{\mathrm{A}}$ is the \textsc{Dec}-\textsc{Acc} trajectory with the smallest $t_{\mathrm{D}}$, by Proposition \ref{propT:2}, we have
\begin{align}
&V(\underline{t},\overline{t})=C^*(\sigma_{\mathrm{A}};t_{k+1},\mathcal{I}_{k+1})\notag\\
&=v_{\mathrm{M}}(t_{0,\mathrm{in}}^{\sigma_{\mathrm{A}}}-t_{k+1})+ \frac{\big(v_{\mathrm{M}}-v_0^{\sigma_{\mathrm{A}}}(t_{0,\mathrm{in}}^{\sigma_{\mathrm{A}}})\big)^2}{2a_{0,\mathrm{M}}}
\end{align}
which is a constant with respect to $\langle\underline{t},\overline{t}\rangle$.

(ii) Similar to (i), $\sigma_{\mathrm{A}}$ also satisfies the robust safety condition on $\mathcal{D}_2$, and we have
\begin{align}
&V(\underline{t},\overline{t})=C^*(\sigma_{\mathrm{A}};t_{k+1},\mathcal{I}_{k+1})\notag\\
&\!=v_{\mathrm{M}}\big[(t_{0,\mathrm{in}}^{\sigma_{\mathrm{A}}}-t_{k+1})-(\overline{t}-\underline{t})\big] +\frac{\big(v_{\mathrm{M}}-v_0^{\sigma_{\mathrm{A}}}(t_{0,\mathrm{in}}^{\sigma_{\mathrm{A}}})\big)^2}{2a_{0,\mathrm{M}}}.
\end{align}
Then the results can be simply checked.

(iii) On $\mathcal{D}_3$, $\sigma_{\mathrm{A}}$ does not satisfy the robust safety condition. By Proposition \ref{propT:2}, the best trajectory is $\sigma_{\mathrm{DA}}(\overline{t})$ if it exists, i.e., the \textsc{Dec}-\textsc{Acc} trajectory satisfying $t_{0,\mathrm{in}}=\overline{t}$. Hence,
\begin{align}
&V(\underline{t},\overline{t})=v_{\mathrm{M}}(\underline{t}-t_{k+1}) +\frac{\big(v_{\mathrm{M}}-v_0^{\sigma_{\mathrm{DA}}(\overline{t})}(\overline{t})\big)^2}{2a_{0,\mathrm{M}}}
\end{align}
which is clearly non-decreasing with $\underline{t}$, and is also non-decreasing with $\overline{t}$ according to Lemma \ref{lemmaT:2}(iii) in Appendix \ref{subsecT:lemma}. Furthermore, note that $\sigma_{\mathrm{DA}}(\overline{t})$ does not exist if and only if $\overline{t}> t_{0,\mathrm{in}}^{\sigma_{\mathrm{D}}}$, where $\sigma_{\mathrm{D}}$ is the \textsc{Dec} trajectory of $s_0$ on $[t_{k+1},+\infty)$, and in this case we have $V(\underline{t},\overline{t})=+\infty$. Therefore, the monotonicity results remain true.
\end{IEEEproof}

\subsection{Proof of Proposition \ref{corT:4} in Section \ref{subsecT:cal}}\label{subsecT:pf_cor4}
\begin{IEEEproof}
(i) The result is direct by Theorem \ref{thmT:1}(i).

(ii) First, it is easy to show that \eqref{eqT:cor4_1} lies in $\mathcal{B}(\mathcal{I}_k)\cap\mathcal{D}_2$ by Assumption \ref{asT:2} and \eqref{eqT:D2}. According to Theorem \ref{thmT:1}(ii), the optimal point in $\mathcal{B}(\mathcal{I}_k)\cap\mathcal{D}_2$ is obtained at the point with minimum $\overline{t}-\underline{t}$. Assume that this minimum occurs at some point $\langle \underline{t}_*,\overline{t}_*\rangle$, and we must have $\overline{t}_*=\underline{\hat{t}_{1,\mathrm{out}}}(\underline{t}_*)$ due to the minimality. Then by Assumption \ref{asT:2}, $\underline{t}_*$ should also be chosen as the smallest value, which results in the point \eqref{eqT:cor4_1}.

(iii) According to Theorem \ref{thmT:1}(iii), we only need to show the point \eqref{eqT:cor4_2} lies in $\mathcal{B}(\mathcal{I}_k)\cap\mathcal{D}_3$. This is also straightforward by Proposition \ref{propT:4a} and \eqref{eqT:D3}.
\end{IEEEproof}

\end{document}